\documentclass[12pt]{spieman}  
\usepackage{amsmath,amsfonts,amssymb}
\usepackage{graphicx}
\usepackage{setspace}
\usepackage{tocloft}
\usepackage{textcomp}

\usepackage{booktabs} 
\usepackage{comment}

\usepackage{array}
\newcolumntype{C}[1]{>{\centering\arraybackslash}p{#1}}

\title{Adaptive optics performance of a simulated coronagraph instrument on a large, segmented space telescope in steady state}

\author[a,*]{Axel~Potier}
\author[a]{Garreth~Ruane}
\author[b]{Chris~Stark}
\author[a]{Pin~Chen}
\author[c]{Ankur~Chopra}
\author[c]{Larry~Dewell}
\author[b,d]{Roser~Juanola-Parramon}
\author[c]{Alison~Nordt}
\author[e]{Laurent~Pueyo}
\author[a]{David~Redding}
\author[a]{A J~Eldorado~Riggs}
\author[f]{Dan~Sirbu}

\affil[a]{Jet Propulsion Laboratory, California Institute of Technology, 4800 Oak Grove Drive, Pasadena, CA 91109}
\affil[b]{NASA Goddard Space Flight Center, 8800 Greenbelt Rd, Greenbelt, MD 20771}
\affil[c]{Lockheed Martin Space, Advanced Technology Center, 3251 Hanover Street, Palo Alto, CA 94304}
\affil[d]{Center for Space Sciences and Technology, University of Maryland, Baltimore County, Baltimore, MD 21250}
\affil[e]{Space Telescope Science Institute, 3700 San Martin Dr, Baltimore, MD 21218}
\affil[f]{NASA Ames Research Center, Space Science \& Astrophysics Branch, Moffett Field, Mountain View, CA 94035}

\cftpagenumbersoff{figure}
\cftpagenumbersoff{table}

\begin{document} 
\maketitle

\begin{abstract}
Directly imaging Earth-like exoplanets (``exoEarths'') with a coronagraph instrument on a space telescope requires a stable wavefront with optical path differences limited to tens of picometers RMS during exposure times of a few hours. While the structural dynamics of a segmented mirror can be directly stabilized with telescope metrology, another possibility is to use a closed-loop wavefront sensing and control system in the coronagraph instrument that operates during the science exposures to actively correct the wavefront and relax the constraints on the stability of the telescope. In this paper, we present simulations of the temporal filtering provided using the example of LUVOIR-A, a 15~m segmented telescope concept. Assuming steady-state aberrations based on a finite element model of the telescope structure, we (1)~optimize the system to minimize the wavefront residuals, (2)~ use an end-to-end numerical propagation model to estimate the residual starlight intensity at the science detector, and (3)~predict the number of exoEarth candidates detected during the mission. We show that telescope dynamic errors of 100~pm~RMS can be reduced down to 30~pm~RMS with a magnitude 0 star, improving the contrast performance by a factor of 15. In scenarios where vibration frequencies are too fast for a system that uses natural guide stars, laser sources can increase the flux at the wavefront sensor to increase the servo-loop frequency and mitigate the high temporal frequency wavefront errors. For example, an external laser with an effective magnitude of -4 allows the wavefront from a telescope with 100~pm~RMS dynamic errors and strong vibrations as fast as 16~Hz to be stabilized with residual errors of 10~pm~RMS thereby increasing the number of detected planets by at least a factor of 4. 
\end{abstract}

{\noindent \footnotesize\textbf{*}Axel Potier,  \linkable{axel.q.potier@jpl.nasa.gov} }
\begin{spacing}{2}   

\section{Introduction}
The 2020 Astronomy and Astrophysics (Astro2020) Decadal Survey report\cite{Astro2020} recommended a large infrared / optical / ultraviolet (IR/O/UV) telescope aiming to obtain spectra of a robust sample of $\sim$25 potentially habitable exoplanets. While the telescope design has yet to be defined for this mission, preliminary modeling has shown that the minimum aperture size to achieve this scientific goal is likely $\sim$6 meters (inscribed) assuming an off-axis, unobscured telescope equipped with a coronagraph instrument\cite{Stark2019}\,. The Large UV/Optical/IR Surveyor (LUVOIR) report\cite{LUVOIR_finalReport} discussed two segmented telescope concepts with coronagraph instruments capable of exceeding these scientific goals: LUVOIR-A and LUVOIR-B, with respective diameters of 15~m (13.5~m, inscribed) and 8~m (6.7~m, inscribed). 

To directly image Earth-sized exoplanets (``exoEarths'') with any telescope design, a coronagraph instrument must provide a stable raw contrast of $\sim$10$^{-10}$, which requires an extremely stable wavefront. Segmented telescopes present unique challenges for high-contrast imaging, such as segment gap obscurations and potentially more dynamic wavefront errors at the most critical mid spatial frequencies. To understand the implications of these issues, NASA's Exoplanet Exploration Program initiated the Segmented Coronagraph Design and Analysis (SCDA) study to assess the viability of imaging exoEarths with segmented space telescopes. To this end, the SCDA modeling team has optimized coronagraph mask designs for segmented telescopes, demonstrated their abilities to reach contrast levels below $10^{-10}$ in simulation\cite{Ruane2017_SPIE,Sirbu2020,Juanola2022}\,, and predicted their scientific yields in an idealized, unaberrated, and stable regime\cite{Stark2019}\,. SCDA also studied the sensitivity of these coronagraphs to optical aberrations and found that a raw contrast level of $\sim$10$^{-10}$ is maintained in the image if the optical path differences caused by segment phasing errors does not exceed $\sim$10~pm~RMS\cite{Pueyo2019,Juanola2022}\,, which is three to four orders of magnitude more stable than the James Webb Space Telescope (JWST)\cite{Lightsey2004,Lightsey2018}\,.

A future segmented flagship mission will have the advantage of JWST heritage to reduce risk and cost of a segmented architecture\cite{Feinberg2016} as well as more recent advances in telescope metrology strategies to reduce motions at the primary mirror level\cite{Bolcar2016, Feinberg2017, Saif2017, Lou2018}\,. In addition, the coronagraph instrument may use a closed-loop wavefront sensing and control (WS\&C) architecture, similar to an adaptive optics (AO) system on a ground-based telescope, to correct dynamic aberrations using the combination of one or more deformable mirrors (DMs) and a wavefront sensor (WFS)\cite{Pueyo2019WP}\,. In this paper, we assess the capability of such a AO system to minimize dynamic aberrations in the coronagraph instrument, stabilize the high-contrast performance in the long-exposure science images, and relax the demanding constraints on the mechanical stability of the telescope. Throughout this paper, we make use of a telescope  finite element model based on the LUVOIR-A design as an informative example, which is the most advanced of its kind prior to the release of the Astro2020 decadal survey report. In future work, these simulations will be adapted to a smaller diameter telescope architecture that is more representative of the flagship mission recommended by the decadal survey. However, the AO performance predictions presented herein highlight the benefits and limitations of a coronagraph instrument with a high-order AO capability on a future space telescope and the developed methods are essential for deriving accurate telescope stability requirements. 

In Section~\ref{sec:Statistics_Wavefront}, we introduce the dynamical aberrations assumed for this study whose spatio-temporal properties are based on a finite element model of a large segmented telescope similar to the LUVOIR-A architecture. The resulting wavefront time series is projected into a modal basis for the purpose of AO correction. In Section~\ref{sec:AOsystem}, we develop an analytical model of an AO system and derive its temporal response. In Section~\ref{sec:OptimizationAndResults}, we optimize the response of the AO system to minimize the wavefront residuals under specific observing circumstances. Then, using an end-to-end forward model of the coronagraph instrument, we predict the closed-loop contrast performance and the exoEarth candidate yield gain provided by the optimized AO system.

\section{Spatio-temporal characteristics of the dynamical wavefront}
\label{sec:Statistics_Wavefront}
\subsection{Structural dynamics of a large segmented telescope}

The remaining wavefront error residuals after a closed-loop adaptive optics system depends on the spatio-temporal properties of the input (or “open-loop”) wavefront aberrations. Lockheed Martin (LM) developed a model of the LUVOIR-A observatory, that integrates the flexible-body dynamics of both the spacecraft bus and the optical payload, a linear optical model of the Optical Telescope Assembly (OTA), control systems for maintenance of payload LOS on the target star, and observatory disturbances that contribute to time-domain variation in the wavefront\cite{Dewell2019}\,. The principal components of this integrated model are described in the following paragraphs. These perturbations are driven by the dynamic interaction of flexible structures with the noise and disturbances of the multi-stage pointing control system.

\paragraph{Spacecraft and payload structural dynamics} The flexible-body dynamics are modeled by a set of linear ordinary differential equations, based on the normal vibration modes computed from a Finite Element Model (FEM) of the respective structures. The second-order vibratory dynamics are diagonalized by the mode shapes, using a set of generalized modal coordinates, and a mode shape matrix converts input physical disturbances (typically forces and torques disturbing the structure) into the modal degrees of freedom, and also converts the generalized coordinates to physical displacement and rotation of important physical location on the structure, such as primary mirror segments. For the modeled LUVOIR-A architecture, the spacecraft mass was 11,490~kg, the payload mass was 23,128~kg; significant structural modes below 250~Hz were extracted from the FEM to capture the dominant structural dynamics.

\paragraph{Vibration isolation and pointing models} To reduce dynamic wavefront errors well below 1 nm RMS and achieve low residual line of sight (LoS) error, the model architecture includes passive isolation between the spacecraft Control Moment Gyroscope (CMG) actuators and surrounding structure, as well as a non-contact pointing stage between the spacecraft and the sensitive science payload. Pointing errors are sensed by the fine guidance mode of LUVOIR’s High Definition Imager and controlled in two stages, by a fast steering mirror feeding back to the Fine Guidance Sensor (FGS) with a 5~Hz closed-loop bandwidth, and the Vibration Isolation and Precision Pointing System (VIPPS) controlling the overall payload rigid-body angular degrees of freedom over a 0.05~Hz bandwidth. VIPPS is a non-contact spacecraft-to-payload architecture whereby payload inertial rigid-body attitude is controlled by voice-coil actuators while payload-spacecraft relative alignment is sensed by non-contact sensors. This sensed interface alignment error is used to compute the spacecraft CMG torque in real-time to maintain positive stroke and gap at the non-contact interface during payload imaging\cite{Pedreiro2003}\,. While no structural contact exists as a load path, residual coupling from spacecraft/payload power and data cables were modeled. The cable stiffness was computed from payload and spacecraft mass properties, such that the translational cable suspension modes were equal to 3~mHz, and the rotational modes were 0.5~mHz.

\paragraph{Linear optical line-of-sight and wavefront error model} Variations in LoS and telescope exit pupil Optical Path Difference (OPD) due to structural dynamic response was modeled as a linear matrix multiplier on the 6-DOF translation and rotation of the 120 hexagonal primary mirror segments, secondary and tertiary mirror, fast steering mirror, and coronagraph instrument. The first-order LoS sensitivity was thus a 2$\times$744 matrix, transforming the 744 displacement degrees of freedom (124$\times$6) into two LoS perturbations. For OPD, the exit pupil was sampled over a 128$\times$128 grid. Thus, the OPD sensitivity was modeled as a 16,384$\times$744 matrix sensitivity, transforming the 744 displacement degrees of freedom into an OPD vector, where each element represents the scalar OPD at a given location in the 128$\times$128 grid.

\paragraph{Modeled disturbances and error sources} The integrated model incorporated disturbances arising from the CMG exported force and torque, VIPPS interface voice coil actuator current drive noise, FGS noise, steering mirror exported loads and servo control error, and VIPPS non-contact sensor noise. Exported loads from CMGs were based on measured harmonic data, with an additive frequency-dependent Model Uncertainty Factor (MUF). Voice coil current drive noise was modeled as additive force noise with specified RMS force, and shaped by a low-pass filter.

\paragraph{}With these physical and parametric disturbance models, three different time series were generated with RMS OPD error of 3.2~pm, 9.9~pm and 114~pm, which we refer to as sample A, B and C, respectively. The three disturbance cases were realized by scaling both the CMG exported loads and the VIPPS voice coil current noise, which were determined to have the highest sensitivity to overall RMS wavefront error. Sample A represents a stable and optimistic scenario, sample B has roughly the level of error specified in the LUVOIR report\cite{LUVOIR_finalReport}\,, while sample C represents a more pessimistic case with $\sim$10$\times$ larger wavefront errors than B. The derived LoS error for all sample B was $<$0.1 milli-arcseconds, which is three times lower than the requirement specified in the LM SLSTD Phase 1 Final Report\cite{LMC2019}\,. While the LoS error is usually dominant in the overall error budget if uncontrolled, we assume that it is negligible because of its relatively small impact on the image contrast. Each data set has a duration of T = 20s, representing the telescope wavefront errors in steady state with residual vibrations. They are sampled with 8000 OPD maps resulting in an effective sample rate of 400 Hz. The feasibility of these levels of stabilization with a 15-m segmented space telescope is beyond the scope of this paper. The telescope structure assumed in our model has not been anchored to test data or optimized for stability. In addition, metrology-based segment position sensing and control models have yet to be incorporated into the integrated model. Therefore, some wavefront patterns contained in the data sets and later presented in this publication may be mitigated through a future modification of the mechanical design or via telescope metrology\cite{Lou2018}\,. For the purposes of this study, we ignore dynamic aberrations at temporal frequencies $\lesssim$1~Hz because we expect those to be mitigated with the aforementioned telescope model improvements and negligible with respect to the high temporal frequency residuals.

\subsection{Modal decomposition of the dynamic errors}
To model the AO performance, we decompose the OPD maps into modes. The full AO system will effectively have independent control loops for each mode running in parallel to maximize efficiency. We use principal component analysis (PCA) to create an orthonormal basis unique to each time series, which maximizes the variance of the phase projected on the first modes\cite{Roddier1999}\,. In this respect, PCA minimizes the mean square error that would result from a modal truncation. Numerically, the PCA is calculated using singular value decomposition (SVD) of the discrete OPD data set. If $X$ stands for the matrix representing the OPD data set, the SVD of $X$ can be written as $X=U\Sigma V^T$, where $\Sigma$ is a rectangular diagonal matrix with the singular values of $X$ sorted in ascending order. $U$ and $V$ are two square matrices whose vectors are orthogonal and called respectively left and right singular vectors. Under this formalism, the principal components are $P=V^T$. Appendix \ref{sec:orthonormal_basis} derives statistical properties of the orthonormal basis relevant to the analysis below.

\begin{figure}[t]
    \centering
	\includegraphics[width=\linewidth]{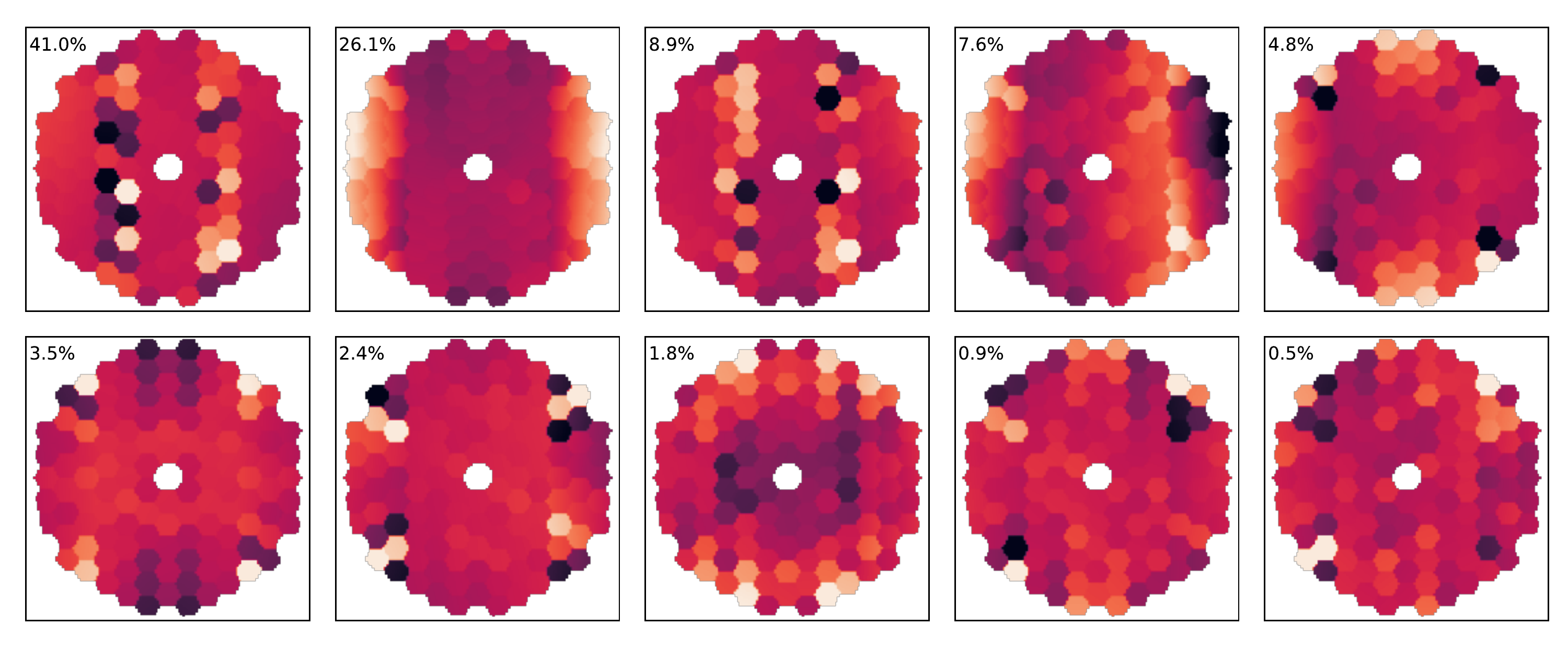}
	\caption{The first ten principal components of the ``sample B" OPD time series ($\sim$10~pm~RMS). The number in the upper left of each panel is the percentage of the total variance represented by each mode.} 
	\label{fig:PCA_modes} 
\end{figure}

Figure~\ref{fig:PCA_modes} shows the first ten principal components of the decomposition of time series B. Here, the first three modes account for 76\% of the total variance and are characterized by a strong vertical shape; thus, the wavefront error is dominated by a small number of structural modes. These effects may be mitigated with a future mechanical optimization of segmented telescope architectures or by advanced telescope metrology. For instance, in our model, a 10x increase in damping of only 5 modes on the payload structure reduces the sample B wavefront error to 2.5~pm RMS. Since the eight first modes represent 96\% of the total variance, we only control these eight modes in our AO performance analysis for sample B, which allows us to simplify subsequent numerical calculations with a minor loss of precision. Except for the second principal component, we note that each mode is dominated by mid-spatial frequencies induced by segment phasing errors. 

\begin{figure}[t]
    \centering
	\includegraphics[width=\linewidth]{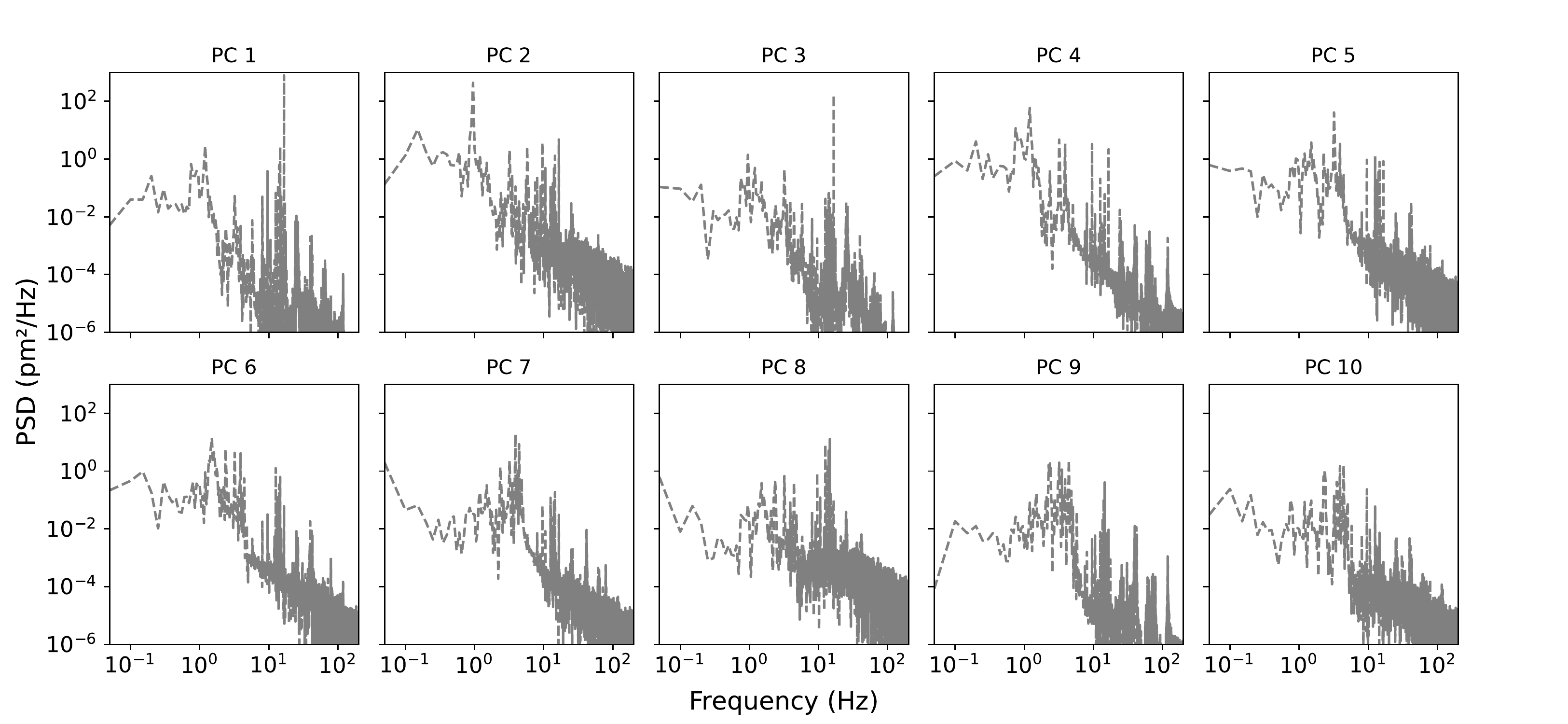}
	\caption{Power spectral densities of the first ten principal components for the ``sample B" OPD time series (i.e. for the spatial modes shown in Fig.~\ref{fig:PCA_modes}).} 
	\label{fig:PCA_PSD} 
\end{figure}

We calculate the temporal power spectral densities (PSDs) for each of the modes in the PCA basis. Figure~\ref{fig:PCA_PSD} shows the result for time series B. The PSDs obey power laws with exponents between -1.5 and -2 for the first ten principal components. Moreover, the variances are dominated by a few localized vibrations at 16.5~Hz in the first and third modes and 0.9~Hz in the second. In fact, the 16.5~Hz vibration accounts for 20\% of the total variance. With this information in mind, there may be considerable room for optimization of the telescope mechanical architecture to improve its dynamic stability by damping this, and other, specific vibration modes. Otherwise, the correction of these vibrations requires an AO control loop bandwidth significantly greater than the vibration frequency.

There are some drawbacks to the basis we have selected. First, the AO system architecture is not taken into account; a better option may be to adapt the basis such that the AO hardware optimally senses and corrects the first modes. For example, the DM may not be able to accurately correct the derived modes because of its finite number of actuators and resulting fitting errors. Additionally, the WFS may not have the resolution to differentiate between the modes accurately. In this paper, we optimize the number of pixels in the WFS to provide the best correction for the derived PCA modes. Secondly, correcting the first PCA modes in order of their OPD variance minimizes the resulting OPD error, but does not necessarily provide the best coronagraph contrast performance. Optimized coronagraphs for LUVOIR are, for instance, more sensitive to mid-spatial aberrations than to low-order aberrations with equal variance. The PCA decomposition might not capture modes with low--but not negligible--variance that have a high impact on the contrast performance of the instrument. In the following sub-section, we carry out contrast calculations using the PCA decomposition and confirm that the modes are mostly ordered by their impact of contrast despite these potential pitfalls. 

\subsection{Contrast calculation}
\label{subsec:Modal_Contrast}
We use a ``compact" end-to-end model of the coronagraph instrument i.e. all the aberrations are collapsed down into the input pupil plane and only Fourier propagation is employed in between optical planes, assuming no particular testbed design nor coronagraph internal aberrations from spectral or polarization errors) to estimate the contrast change due to each OPD error map. The numerical simulation uses the PROPER\cite{Krist2007} and FALCO\cite{Riggs2018} software packages to generate images for each of the 8000 OPD time steps in a 20\% spectral bandwidth centered at 550nm using LUVOIR's narrow field-of-view binary Apodized Pupil Lyot Coronagraph (APLC)\cite{Pueyo2019, Juanola2022}\,. The APLC attenuates the starlight with a binary apodizer optimized with respect to the telescope architecture, inner working angle and spectral bandwidth that is introduced at a relayed telescope pupil, a focal plane mask (here of radius 26.5~mas) to occult the core of the stellar Point Spread Function (PSF) associated with a Lyot stop  in the following pupil plane whose inner and outer diameters are 12\% and 98.2\%, respectively, of the pupil diameter. The resulting region where the starlight is attenuated, or ``dark hole" (DH) is an annulus between 3.5 and 12.0~$\lambda/D$ from the star. For time series B, we find that the mean intensity, normalized by the maximum of the unocculted PSF, is equal to $9.5\times10^{-11}$ in the DH. Comparing this result with the normalized intensity induced by sample B that is reduced to its first 10 principal components, the mean difference between the normalized intensity at each time steps is $4.2\times10^{-13}$ while the standard deviation of the differences is $8.8\times10^{-13}$. Considering only the first 10 principal components therefore induces $<$1\% error in the contrast calculation.

\begin{figure}[t]
    \centering
	\includegraphics[width=\linewidth]{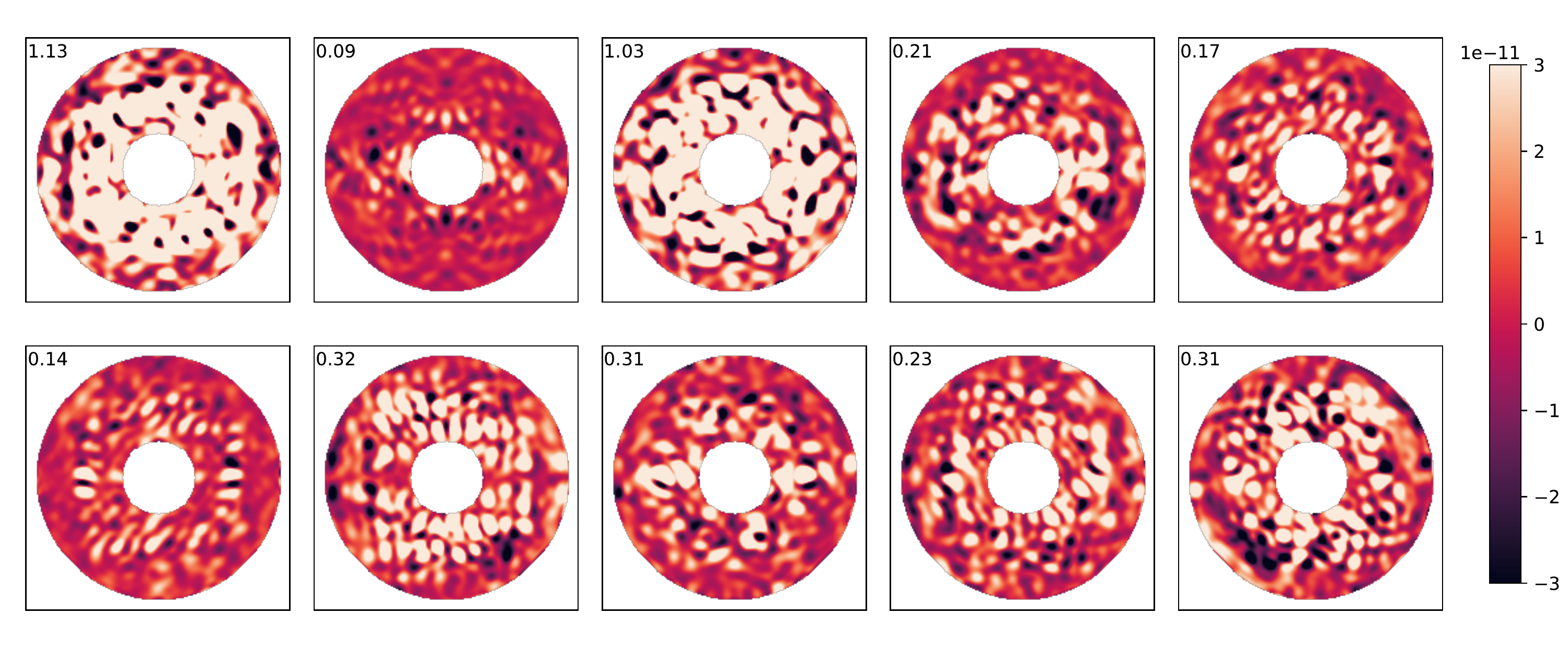}
	\caption{Contrast change in the image plane due to 10~pm RMS of each principal component (ordered one to ten from left to right). The RMS normalized intensity calculated in the dark hole is also displayed in the upper left of each panel ($\times 10^{-10}$).} 
	\label{fig:PCA_intensity} 
\end{figure}
\begin{figure}[t]
    \centering
	\includegraphics[width=8cm]{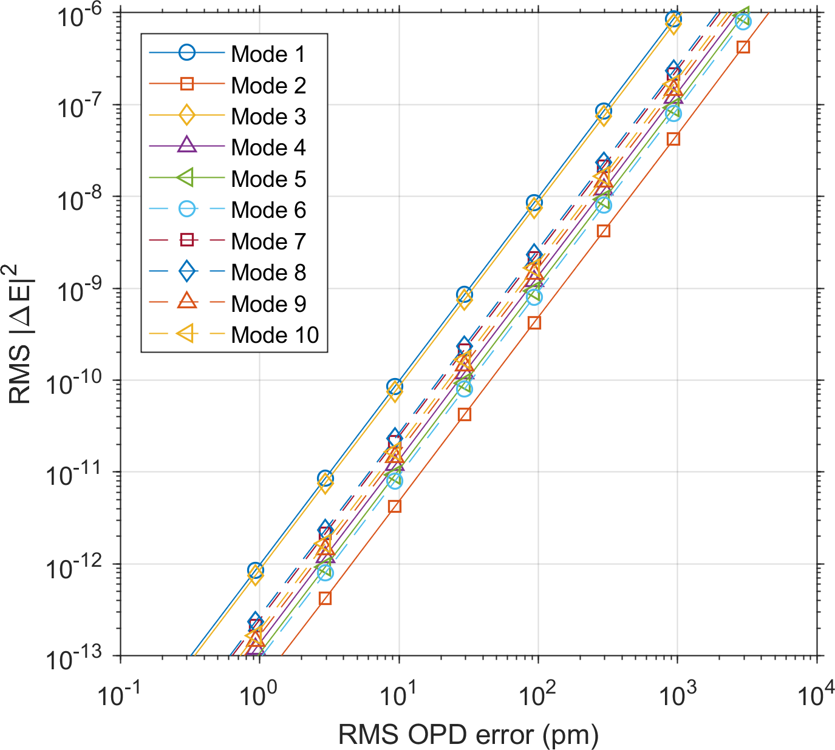}
	\caption{Normalized intensity induced by each of the first ten PCA modes with respect to the level of aberration at the central wavelength.} 
	\label{fig:DeltaE_vs_OPD} 
\end{figure}

In Fig.~\ref{fig:PCA_intensity}, we show the change in contrast induced by the first individual PCA modes when the spatial variance at coronagraph entrance pupil is equal to 10~pm RMS. Figure~\ref{fig:DeltaE_vs_OPD} shows the mean normalized intensity $|\Delta E_i|^2$ induced by each individual mode $i$ at the science detector versus the level of OPD error for the first 10 modes. The simple quadratic relationship allows us to easily estimate the contrast impact of the OPD data sets by scaling to any amount of OPD error. In both Figs.~\ref{fig:PCA_intensity} and \ref{fig:DeltaE_vs_OPD}, it is evident that the contrast is most sensitive to the first and third principal components. Indeed, 10~pm RMS of these modes individually bring the mean contrast in the DH above $10^{-10}$. Conversely, the second principal component has a relatively small effect on the contrast. In time series B, the order of magnitude of the contrast induced by the second component ($2.3\times10^{-12}$) is similar to the contrast induced by the fourth ($1.6\times10^{-12}$) despite the $>$3$\times$ larger OPD variance in the second mode. For the other components, the mean contrast values are roughly the same for a constant OPD variance. This justifies using the PCA decomposition because the contrast impact of each mode in the time series (for the most part) decreases with the mode number and the contrast error due to truncation is negligible. In the remainder of this publication, the long-exposure post-AO contrast performance is calculated using Eq.~\ref{eq:MeanContrast_calculation} in appendix~\ref{sec:long_exp_contrast}. 

Table~\ref{table:modeParams} summarizes the OPD error by mode for each time series and the relative impact of each mode on the contrast. The impact of each mode on contrast is roughly proportional to the product of the last two columns. See Appendix \ref{sec:timeseries_properties} for more detail on the properties of time series A and C.

\begin{table}[htp]
	\caption{Modal decomposition properties of each time series. The first column is the mode number. For each time series, the three columns are (1)~the RMS OPD error in units of pm, (2)~the percentage of the OPD variance explained by each mode, and (3)~the RMS $|\Delta E|^2$ at the central wavelength for a 10~pm~RMS wavefront error scaled by $10^{10}$.}
	\centering
	\begin{tabular}{|c|*{3}{C{1cm}}|*{3}{C{1cm}}|*{3}{C{1cm}}|}
        \toprule
       PCA Mode & \multicolumn{3}{c|}{Time series A} & \multicolumn{3}{c|}{Time series B} & \multicolumn{3}{c|}{Time series C}\\
		\midrule
         & OPD & \%Var & $|\Delta E|^2$ & OPD & \%Var & $|\Delta E|^2$ & OPD & \%Var & $|\Delta E|^2$ \\
		\midrule
		1 & 2.0 & 39.8 & 0.1 & 6.3 & 41.0 & 1.0 & 73.7 & 41.9 & 0.2\\
        \midrule
        2 & 1.4 & 18.9 & 0.2 & 5.1 & 26.1 & 0.0 & 59.2 & 27.0 & 0.9\\
        \midrule
        3 & 0.9 & 7.9 & 0.1 & 2.9 & 8.9 & 0.9 & 29.4 & 6.7 & 0.7\\
        \midrule
        4 & 0.9 & 7.1 & 0.2 & 2.7 & 7.6 & 0.1 & 28.9 & 6.4 & 0.4\\
        \midrule
        5 & 0.8 & 6.7 & 0.3 & 2.2 & 4.8 & 0.1 & 25.2 & 4.9 & 0.1\\
        \midrule
        6 & 0.6 & 3.0 & 0.2 & 1.9 & 3.5 & 0.1 & 21.2 & 3.5 & 0.1\\
        \midrule
        7 & 0.5 & 2.8 & 0.9 & 1.5 & 2.4 & 0.2 & 18.7 & 2.7 & 0.2\\
        \midrule
        8 & 0.4 & 1.9 & 0.3 & 1.3 & 1.8 & 0.3 & 16.7 & 2.1 & 0.2\\
        \midrule
        9 & 0.4 & 1.7 & 0.4 & 0.9 & 0.9 & 0.2 & 12.4 & 1.2 & 0.2\\
        \midrule
        10 & 0.4 & 1.4 & 0.2 & 0.7 & 0.5 & 0.2 & 10.3 & 0.8 & 0.2\\
        \midrule
		\bottomrule
	\end{tabular}
	\label{table:modeParams}
\end{table}

\section{Analytical modeling of an adaptive optics system}
\label{sec:AOsystem}
\subsection{Wavefront Sensing and Control architecture}
In this study, we first assume that static aberrations are compensated before the science operations using conventional focal plane wavefront sensing and control techniques\cite{GiveOn2007SPIE,GiveOn2011}\,. The intensity in the DH is therefore at its minimum and the raw contrast is only limited by the coronagraph's diffraction pattern. We then introduce dynamical telescope aberrations that occur during the science operations and cause stellar leakage in the DH. To correct these aberrations, we need to use a wavefront sensor that (1)~operates simultaneously with science observations, (2)~has a sufficient spatial resolution to sense for segment phasing errors, (3)~is sensitive to picometer level phase aberrations, and (4)~is in a common beam path with the coronagraph. We therefore consider the example of a Zernike Wavefront Sensor (ZWFS) built in to the coronagraph's focal plane mask (FPM), which reflects out-of-band light using a dichroic on the FPM substrate (while part of the in-band light can also be reflected by an occulting spot on the FPM). This type of in-situ ZWFS design has been shown to be sensitive to picometer-level wavefront variations in the laboratory\cite{Ruane2020}\,. For the sake of simplicity, our simulation effectively assumes an ideal in-band ZWFS that collects all of the light passing through the 20\% bandwidth visible filter. In reality, the ZWFS may operate in a different band than science observation's filter, but this distinction is not important to our analysis. In accordance with the small level of aberrations expected, we consider the relationship between measured intensity in the ZWFS and the phase to be linear during operations. The sensitivity of a ZWFS is often parameterized by $\beta$, which relates the phase error to the photon noise\cite{Guyon2005}\,. Here, we round $\beta = 1$ at each wavelength and spatial frequency in a Fourier basis, while $\beta$ has recently been derived more precisely considering the realistic diffraction from the focal plane mask\cite{Ruane2020}\,. We ignore any artifact or bias originating from the WFS detector.
Our assumed telescope and WFS parameters are summarized in Table~\ref{table:TelescopeParameters}. For the correction, we ignore fitting errors due to the finite number of DM actuators and confirmed that the residuals are negligible for a 64$\times$64 actuator DM and the OPD modes derived above. The DM that is conjugated to the telescope pupil is used to correct for both static aberrations and the dynamical aberrations sensed with the ZWFS.

\begin{table}[htp]
	\caption{Telescope Specifications used to analyze the AO system performance. The throughput is the optical transmission from the telescope primary mirror to the wavefront sensor (quantum efficiency included). The detector read out noise (RON) is assumed to be 1 electron per pixel which is conservative with respect to state-of-the-art detectors in AO ($\sim0.3e^-$ for the EMCCD) and the arisen of more advanced detectors like MKIDs\cite{Mazin2012}\,. This value multiplies by 10 the RON/photon noise ratio in Eq.~\ref{eq:SNR-1} with respect to the 0.3$e^-$ value adopted in Douglas et al.}
	\centering
	\begin{tabular}{lcc}
		\toprule
		Inscribed diameter & $D$ & 14~m \\
		\midrule
		 Throughput & $\mathcal{T}$  & 0.3 \\
		 \midrule
		 Central wavelength & $\lambda$  & 545~nm \\
		 \midrule
		 Spectral bandwidth & $\Delta\lambda/\lambda$ & 20\% \\
		 \midrule
		 WFS detector RON & $\sigma_{RON}$ & 1$e^-$ \\
		 \midrule
		 Vega photon flux & $\Phi_{Vega,ph}$ & 10$^8$ m$^{-2}$s$^{-1}$nm$^{-1}$\\
		\bottomrule
	\end{tabular}
	\label{table:TelescopeParameters}
\end{table}

\subsection{System frequency response}
To stabilize the wavefront variation, a conventional linear AO system acts like a temporal filter on the modal coefficients at a given time $a_i(t)$. The corrected coefficients, $a^\prime_i(t)$, are given by the convolution between the impulse response and the input modal coefficient in the time domain:
\begin{equation}
\label{eq:input_output}
a^\prime_i(t)=h_i(t)\ast a_i(t) \quad ,
\end{equation}
where $h_i(t)$ is the AO impulse response function. 
According to the convolution theorem, the Fourier transform of a convolution of two signals is the product of their Fourier transforms. Then, the square of the Fourier transform of Eq.~\ref{eq:input_output} gives the PSD:
\begin{equation}
\label{eq:TFdefinition}
PSD^\prime_i(f)=|H_i(f)|^2\cdot PSD_i(f) \quad ,
\end{equation}
where $H_i(f)$ is the AO transfer function of the $i^{th}$ mode and $f$ is the temporal frequency. However, the AO system also introduces noise during the correction process. The resulting PSD of each mode is a function of the input wavefront error and noise:
\begin{equation}
\label{eq:PSDcorrection}
PSD^\prime_i(f)=|H_i(f)|^2\cdot PSD_i(f)+|H_{N,i}|^2\cdot N_i(f) \quad ,
\end{equation}
where $H_{N,i}(f)$ and $N_i$ stands respectively for noise transfer function and the squared Fourier transform of the noise distribution introduced when sensing the $i^{th}$ mode.

\begin{figure}[t]
    \centering
	\includegraphics[width=0.8\linewidth]{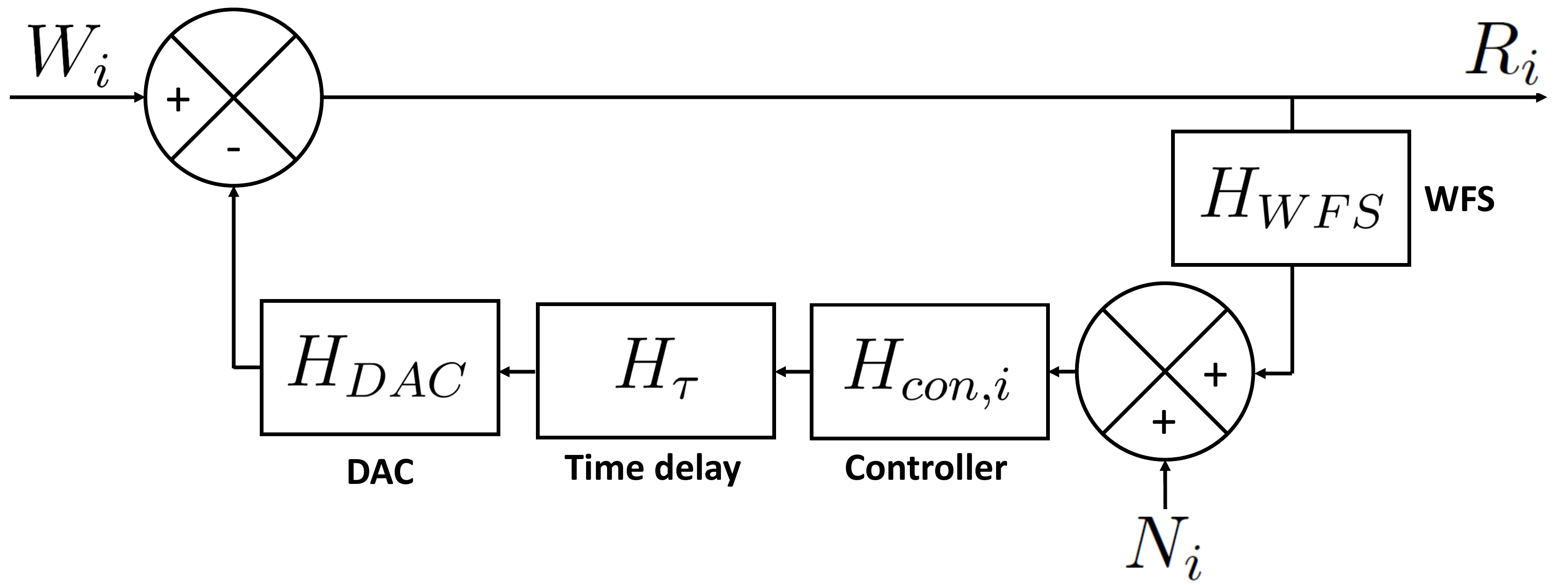}
	\caption{Block diagram of an AO control loop for a single mode in the case of a simple integrator.} 
	\label{fig:AOloop} 
\end{figure}

The block diagram in Fig.~\ref{fig:AOloop} represents the AO temporal model assumed for this study. The total temporal response of the system includes contributions from the wavefront sensor, the controller, the servo-lag error, and the digital-to-analog converter (DAC) which applies the signal to the DM. The transfer function of each individual subsystem is derived in appendix~\ref{sec:AO_loop_TF}. The DM transfer function is not taken into account here but could be readily included in future studies. According to this scheme, the wavefront residuals are given by
\begin{equation}
R_i=W_i-H_{WFS} H_{con,i} H_\tau H_{DAC} R_i\quad , 
\end{equation}
where $R_i$ and $W_i$ are the Fourier transforms of the corrected (residual) and input wavefronts, respectively. We can therefore derive the rejection transfer function as:
\begin{equation}
\label{eq:correction_TF}
H_i=\frac{1}{1+H_{WFS} H_{con,i} H_\tau H_{DAC}}\quad .
\end{equation}

We also derive an analytic expression for the noise transfer function only considering wavefront sensor noise (i.e. electronic read out noise) and photon noise injected before the controller as shown in Fig.~\ref{fig:AOloop}. The noise transfer function relates the residual signal introduced by the system in the presence of only noise. In that case, 
\begin{equation}
R_i= - H_{con,i} H_\tau H_{DAC} ( N_i + H_{WFS} R_i ) \quad .
\end{equation}
and noise transfer function is
\begin{equation}
\label{eq:noise_TF}
H_{N,i}= - \frac{H_{con,i} H_\tau H_{DAC}}{1+H_{WFS} H_{con,i} H_\tau H_{DAC}} \quad .
\end{equation}

\subsection{Optimization of the temporal response}
\subsubsection{Optimized integrator}
\begin{figure}[t]
    \centering
	\includegraphics[width=0.8\linewidth]{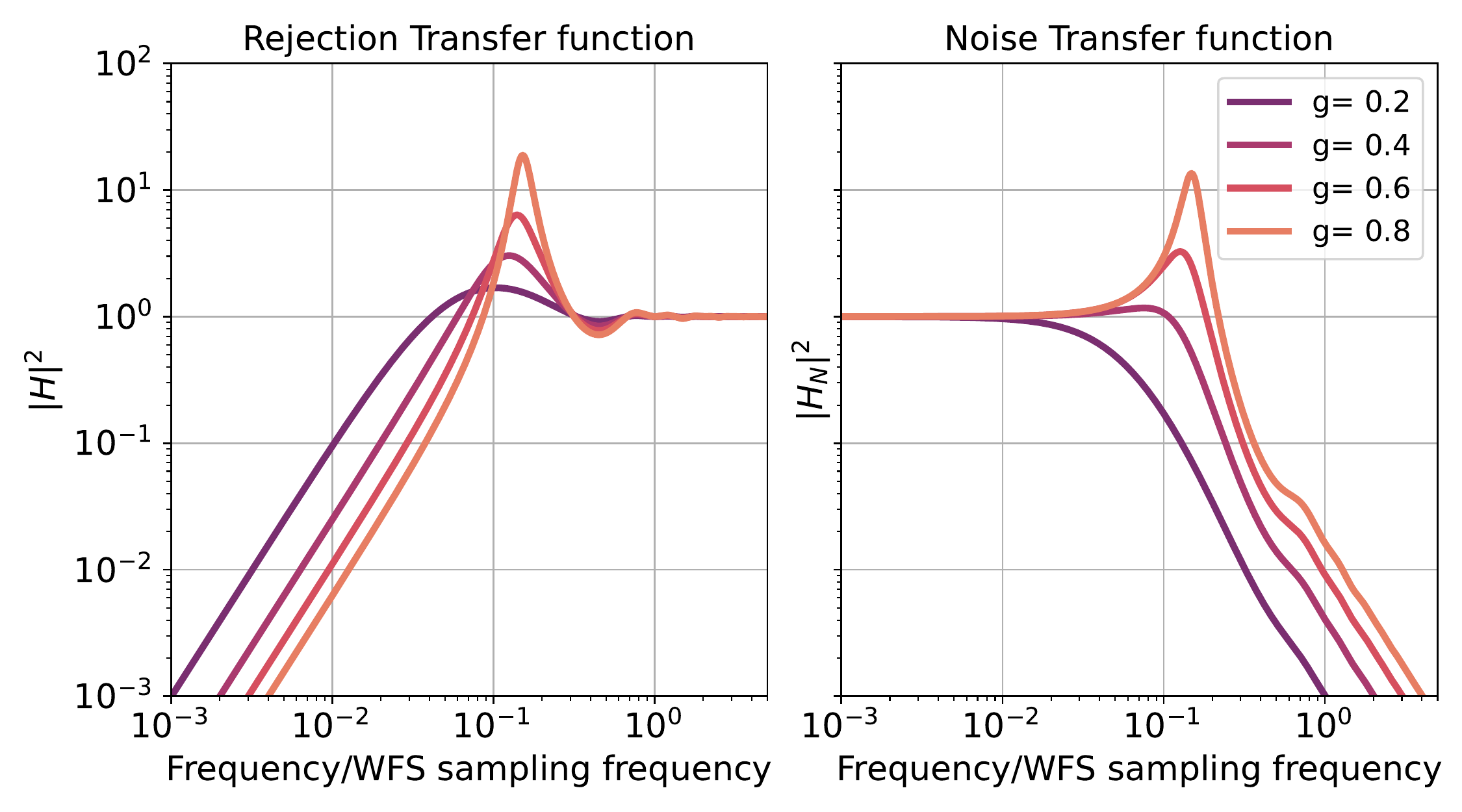}
	\caption{Squared modulus of the rejection transfer function (left panel) and noise transfer function (right panel) with varying AO gain. The sampling frequency is 100~Hz and the pure delay is equal to 1 frame.} 
	\label{fig:TransferFunction} 
\end{figure}

Figure~\ref{fig:TransferFunction} shows the squared modulus of the rejection and noise transfer functions with respect to the gain $g_i$ in the case of a simple integrator (see Eq.~\ref{eq:TF_integrator} in appendix~\ref{sec:AO_loop_TF}) assuming a pure delay of 1 frame. On one hand, the rejection transfer function shows that the telescope stability requirements can be relaxed for frequencies $\lesssim1/10$ of the WFS sampling frequency. For example, assuming a 100~Hz control loop and zero noise, an original telescope stability requirement of 10~pm RMS for a 1~Hz disturbance would be relaxed to 30~pm RMS using an integrator gain of 0.2 and to 100~pm~RMS for a gain of 0.6. The AO cutoff bandwidth (i.e., the maximum frequency where $|H_i|^2 < 1$) also increases with the gain. On the other hand, the signal is amplified for temporal frequencies higher than the AO bandwidth and the amplification increases with the gain. Moreover, the integral of the noise transfer function also increases with the gain. Therefore, the gain needs to be optimized to balance the correction of the incoming wavefront and the propagation of noise in the system as well as avoiding any substantial amplification of wavefront variations that are too fast for the AO system to correct. The gain should be optimized separately for each mode, depending on both the modal PSD and the noise amplitude, in order to get the minimal residual:
\begin{equation}
\label{eq:optimize_SI}
g_i=\text{argmin}\,\sigma_{T,i}^2(g_i,SNR) \quad .
\end{equation}

\subsubsection{Predictive controller}
The correction provided by an integrator is limited by the AO bandwidth and, as a result, faster vibrations will not be corrected by AO system. A potential solution is to replace the integrator with a predictive control algorithm. Indeed, such methods are able to partially mitigate some localized vibrations in the perturbation's PSD that are faster than the integrator AO bandwidth. However Bode's theorem imposes restrictions on the shape of the rejection transfer function \cite{Guesalaga2012}\,. Indeed, the overall integral of the rejection transfer function is fixed by the sampling frequencies and hardware delays. It follows that the perturbation rejection tends to be worse in the vicinity of the targeted vibration frequency with respect to the simple integrator case. The predictive controller also has to be shaped according to the open-loop PSD. Prior knowledge of the perturbation is therefore required to optimize the controller coefficients (see Eq.~\ref{eq:Predictive_controller} in appendix~\ref{sec:AO_loop_TF}) and minimize the residuals:
\begin{equation}
\label{eq:optimize_PC}
(a_0,...,a_p,b_1,...,b_q)_i=\text{argmin}\,\sigma_{T,i}^2(a_0,...,a_p,b_1,...,b_q)_i \quad.
\end{equation}
This PSD knowledge requirement limits the performance of predictive control on ground-based facilities\cite{vanKooten2019}\,. For example, observing conditions can quickly change subject to properties of the air turbulence in the atmosphere and telescope dome. As a result, predictive control may require a reevaluation of the perturbation's PSD that is too frequent to be conveniently applied. On the other hand, space-based telescope might offer a steady state where the coefficients may not require regular re-optimization. The wavefront errors would therefore be described by consistent and potentially well-characterized spatial modes and temporal PSDs.  

Both predictive control and a simple integrator require some knowledge of the perturbations to minimize the residuals. 
In practice, minimizing the integral in Eq.~\ref{eq:PSDcorrection} is equivalent to minimizing:
\begin{equation}
\int |H_i(f)|^2\cdot PSD_{meas,i}(f) \, df \quad ,
\end{equation}
where $PSD_{meas,i}$ is the PSD measured by the AO wavefront sensor in open loop\cite{Dessenne1998}\,. The corresponding time series can therefore be measured directly and the parameters of the controller can be optimized based on those measurements. However, we use the full Eq.~\ref{eq:PSDcorrection} in this report since we already have prior knowledge of the OPD time series of aberrations.

\section{Performance improvement provided by AO}
\label{sec:OptimizationAndResults}
\subsection{Residual variance and stellar magnitude}
In this section, we calculate the closed-loop AO residuals assuming the incoming perturbations described in Sec.~\ref{sec:Statistics_Wavefront}. We have shown that the performance depends on the SNR of the WFS measurement and thus on the magnitude of the AO guide star. We therefore optimized the controller gain as a function of the stellar magnitude between -4 and 11 to cover the broad stellar sample targeted by LUVOIR-A\cite{LUVOIR_finalReport}\,. The resulting gains are used to minimize the residual PSD of each independent mode derived in Sec.~\ref{sec:Statistics_Wavefront}. The noise PSDs of such modes are modeled using the method described in appendix \ref{sec:WFSphotonnoise}. The AO rejection and noise transfer functions are calculated through Eq.~\ref{eq:correction_TF} and Eq.~\ref{eq:noise_TF} respectively.

\subsubsection{Optimized integrator}
\begin{figure}[t]
    \centering
	\includegraphics[width=\linewidth]{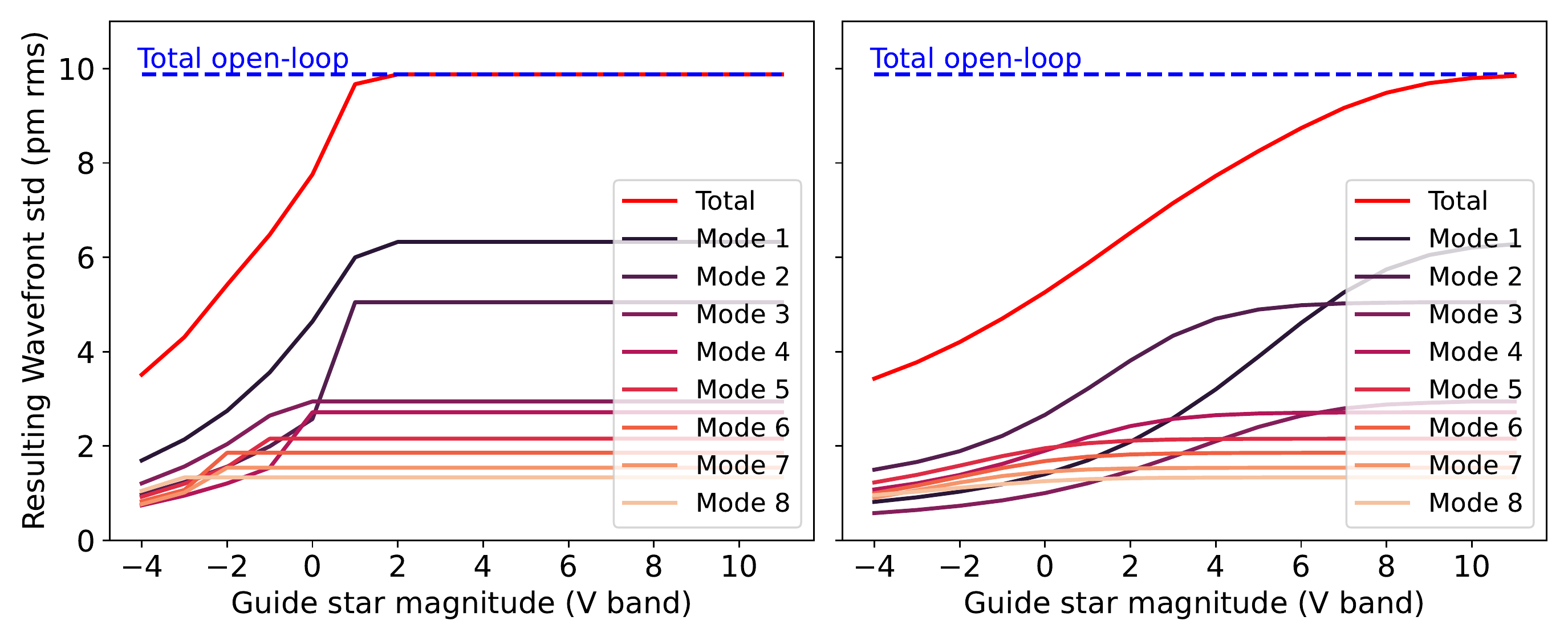}
	\caption{Left: Standard deviation of the residual wavefront after a conventional AO correction (optimized integrator) running at 1~kHz. Right: Same plot using predictive control at 100~Hz. The input dynamical aberrations is time series B (total open-loop RMS OPD error of 9.9~pm~RMS). The residual of the eight first principal components is also shown. These plots demonstrate the optimized integrator can be used to minimize the total wavefront errors if $M_V<2$ while predictive control can work on stars as faint as $M_V<8$. } 
	\label{fig:Variance_Magnitude} 
\end{figure}
We first assume the controller is a simple integrator, the servo-loop frequency sampling is 1~kHz, and the WFS is sampled with 64 pixels across the beam. We also assume 1 frame of pure delay. The other relevant parameters are given in Table~\ref{table:TelescopeParameters}. The gain is optimized for each mode via a Broyden-Fletcher-Goldfarb-Shanno (BFGS) algorithm. For a wavefront error following the statistics of time series B, we plot the residual wavefront standard deviation as well as each mode residual standard deviation with respect to the stellar magnitude in Fig.~\ref{fig:Variance_Magnitude}. For an initial wavefront variation of about 10~pm RMS, the AO system brings an improvement when the visible magnitude of the observed star is $<$1 (i.e. brighter than 1st magnitude). Above this magnitude (i.e. fainter than 1st magnitude), the modal noise $N_i$ is too high for the AO system to correct any modes and all the modal gain are set to 0. Between $M_V=0.2$ and $M_V=1$, only the first principal component is controlled by the AO system. Indeed, the ratio between $PSD_i$ and $N_i$ in Eq.~\ref{eq:PSDcorrection} is the highest for this mode owing to the PCA decomposition. This means that, for the assumed pixel sampling and exposure time, no other mode or decomposition can be corrected for stellar magnitude above 1. A longer exposure time could be used to correct slow varying drifts when observing stars at higher magnitudes. However, most of the OPD variance would remain uncorrected and the AO system would not bring any significant modifications in the OPD statistics. This result implies that a laser guide star (LGS) may be required to stabilize all the mid-order spatial modes under 10~pm RMS if the AO controller is an integrator. Figure~\ref{fig:Variance_Magnitude} also shows that the total standard deviation would be divided by two for $M_V=-2.3$, which is brighter than a typical star in a direct imaging survey.

\subsubsection{Predictive control}
\begin{figure}[t]
    \centering
	\includegraphics[width=\linewidth]{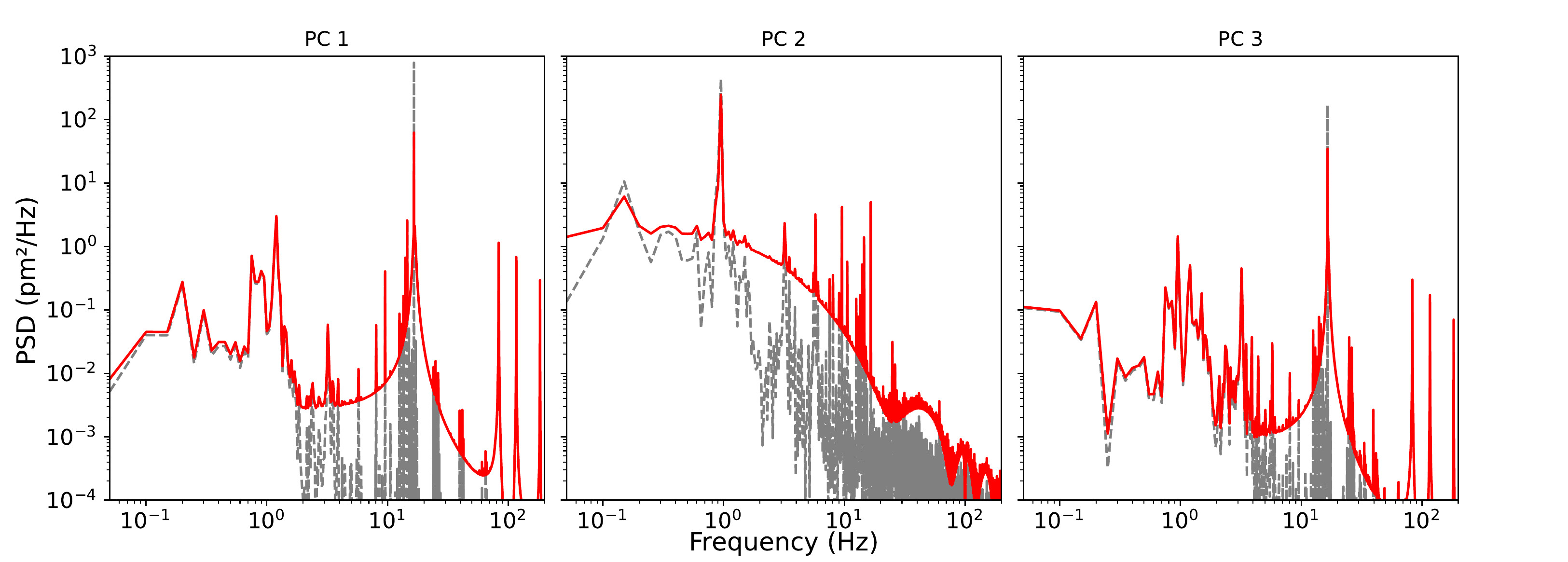}
	\caption{PSDs of the first three principal components of time series B before (grey) and after (red) correction with an AO system running at 100~Hz with a predictive controller. The wavefront sensor is sampled at 64~pixels across the beam and the magnitude of the target star is equal to 3 in V-band. As desired, the predictive controller reduces the highest magnitude peaks of the PSD, thereby reducing the dynamical wavefront error substantially.} 
	\label{fig:PCA_PSD_PC} 
\end{figure}

The numerical simulations above assumed an optimized integrator. We expect better performance with predictive control since it should be able to focus on mitigating vibrations (e.g. at 0.9~Hz and 16.5~Hz in time series B). Figure~\ref{fig:PCA_PSD_PC} shows the PSD of the first three modes before and after correction with predictive control for $M_V=3$. The loop is assumed running at 100~Hz in order to increase the signal-to-noise ratio by $\sqrt{10}$ with respect to the simulation in Fig~\ref{fig:Variance_Magnitude}. We choose $p=q=2$ in Eq.~\ref{eq:Predictive_controller} which implies that we optimize five different parameters in Eq.~\ref{eq:optimize_PC}. We use a constrained optimization method to minimize the cost function in the frequency domain and the parameters are all bounded in between -1 and 1, except $a_0$ whose value is between 0 and 1. While these bounds and algorithms have shown reliable results, the use of recursive algorithms applied in the time domain such as Kalman estimators \cite{Leroux2004,Kulcsar2006,Poyneer2010} whose parameters are optimized by solving a Riccati's equation \cite{Guesalaga2012,Males2018b} may be required to ensure the stability and robustness of the correction. Their implementation is beyond the scope of this work.

Predictive control helps correct one specific vibration for each mode in Fig.~\ref{fig:PCA_PSD_PC}. To do so, the signal is also amplified at other frequencies, as expected from Bode's theorem, which limits further mitigation of these modes. However, the variance residuals is minimized overall. The right panel of Fig.~\ref{fig:Variance_Magnitude} shows the standard deviation of the residuals with respect to the observed star magnitude. Predictive control is able to minimize the wavefront variance with stars brighter than $M_V=10$ which constitutes the majority of the natural stars that will be targeted by a future direct imaging mission. This represents an important improvement with respect to the optimized integrator.

\subsection{Contrast predictions}
\label{subsec:Contrast_Performance}
\begin{figure}[t]
    \centering
	\includegraphics[width=10cm]{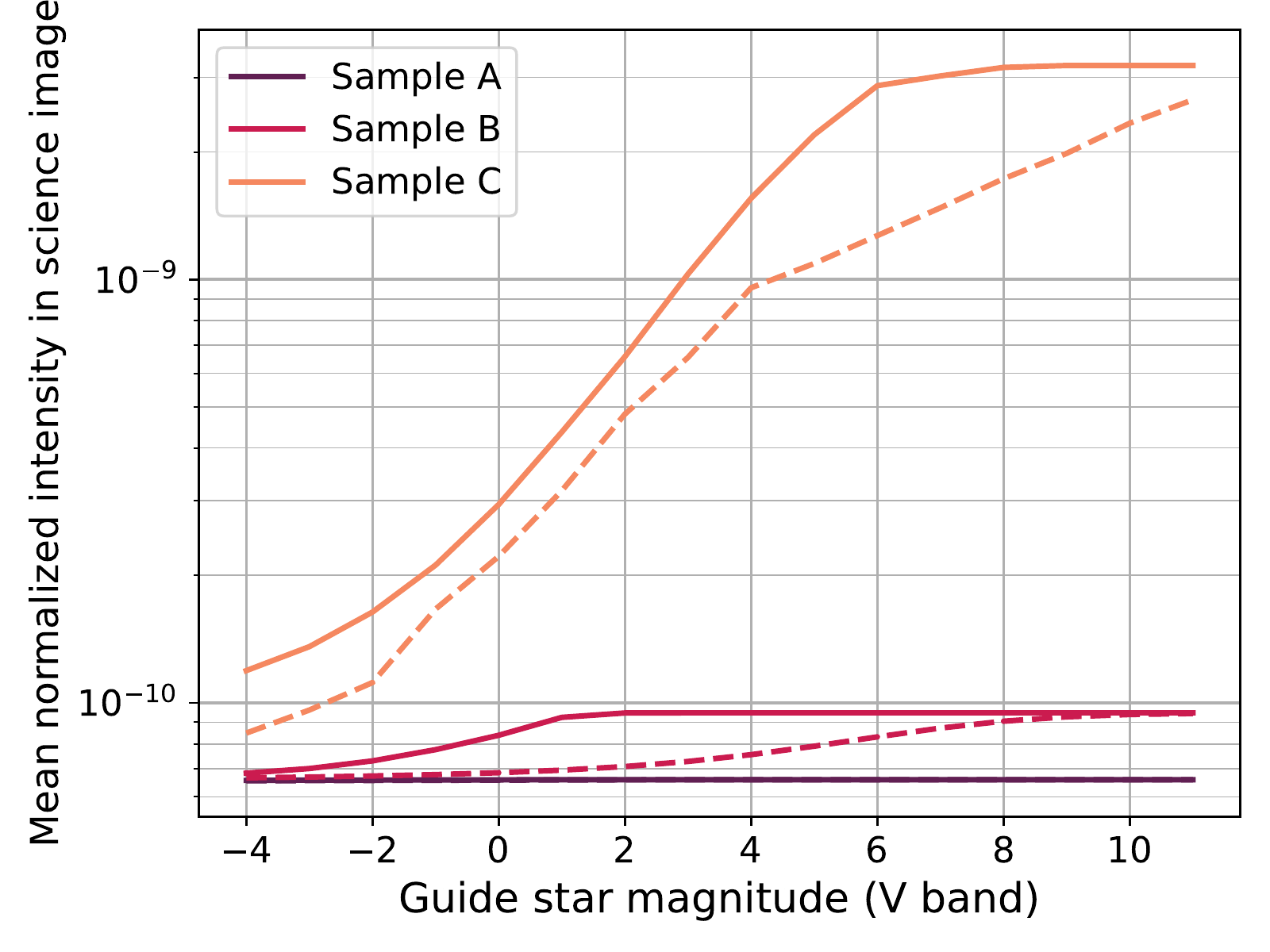}
	\caption{Mean normalized intensity in the dark hole after the wavefront being corrected by an AO system. The AO performance depends on the magnitude of the observed star and on the initial level of turbulence. Purple, pink and orange curves use respectively samples A, B and C as inputs of the simulations. Continuous lines represent results for the optimized integrator while the dashed lines are for predictive control.} 
	\label{fig:Contrast_vs_magnitude} 
\end{figure}

We reproduced an equivalent analysis for time series A and C where we find a similar but not identical PCA decomposition (see Appendix \ref{sec:timeseries_properties} for more detail on the time series statistics). The modes are then corrected independently with optimized AO systems. For the three time series samples, Fig.~\ref{fig:Contrast_vs_magnitude} shows the normalized intensity in the DH annulus of the narrow field-of-view APLC coronagraph, calculated by Eq.~\ref{eq:MeanContrast_calculation}, versus the magnitude of the AO guide star. Here, the residuals for samples A, B and C are all calculated using an optimized integrator running at 1~kHz. We also present the normalized intensity when the time series B and C are corrected with predictive control. The best contrast result between a servo-loop frequency of 100~Hz and 1~kHz is adopted at each magnitude. 

First, Fig.~\ref{fig:Contrast_vs_magnitude} shows that AO is not required in the best case scenario (time series A). Indeed, the wavefront is already low enough before any correction that the normalized intensity is dominated by the static diffraction pattern of the APLC. Sample B's scenario is quite different. If the overall contrast budget for the telescope is on the order of $10^{-10}$, minimizing dynamical aberrations is required because the normalized intensity due to the dynamics alone is $9.45\times10^{-11}$ for sample B in open loop. However, a simple integrator would require a LGS to achieve a significant correction. A $M_V=-4$ star, for instance, would improve the contrast down to sample~A's contrast performance. The coronagraph performance also benefits from predictive control by mitigating the strong vibrations in the first and third principal components. For $M_V<8$, the variance of these two modes is reduced which decreases the dynamical aberrations significantly. Finally, in the worst case scenario where the wavefront is above 100~pm RMS (sample C), the AO system is able to correct aberrations with natural guide stars for either controller. However, predictive control provides better contrast results. Again, there are two different regimes. For natural guide stars, the AO loop should run at $\sim$100~Hz because higher frequencies would make the WFS photon starved leading to insufficient sensitivity. For brighter stars, or LGS, the correction remains limited at such frequencies since the pm level sensitivity has been already reached while high frequency vibrations remain uncorrected. Decreasing the WFS exposure time to correct for these vibrations when observing brighter stars improves the performance. With such an optimization and the essential contribution of a LGS, it is possible to achieve a normalized intensity of $10^{-10}$ in this 100~pm RMS scenario. For the most cost-effective approach, it may be optimal to avoid launching additional spacecrafts for the LGS and consider strategies that minimize the overall dynamical aberrations levels at the telescope level to a few tens of pm RMS and take advantage of the internal AO system to correct the residual wavefront error within the coronagraph instrument.

\subsection{Yield improvement provided by AO}
\label{subsec:yield_performance}

A useful performance metric for a high-contrast imaging mission is the overall exoplanet yield since it takes into account the properties of the target sample including stellar type and diameter, as well as the instrument properties and a given resource budget (mission time, spacecraft fuel). The yield of potentially Earth-like planets, or exoEarth candidates (EECs), in particular has been used extensively to compare different flagship mission concepts proposed as part of the Astro2020 decadal survey. Here, the yield is defined as the expected number of detected EEC at $SNR>7$ in V-band. We require each system to be observed a minimum of six times for planet detection and orbit determination and budget for spectral characterization observations to search for water vapor at 950~nm on each detected EEC (spectral resolution $R=70, SNR=10$). In this section, we use the Altruistic Yield Optimization\cite{Stark2014} (AYO) to understand the impact of the AO system on the estimated EEC yield of a direct imaging mission that makes use of a large segmented telescope. We make identical assumptions to those listed in previous AYO studies\cite{Stark2019}\,. Briefly, we use the same definition of EECs as the LUVOIR and HabEx Study final reports\cite{LUVOIR_finalReport,HabEx_finalReport}\,, adopt a self-consistent occurrence rate of $\eta_\text{Earth}=0.24$, and assign all EECs an wavelength-independent geometric albedo of 0.2. We model a total of 2 years of science time including overheads while setting the maximum single observation time to two months. AYO then optimizes the target selection, exposure time, and cadence of all observations to maximize the EEC yield given our instrument models.

For simplicity, we assume a single guide star magnitude for the whole mission. This is equivalent to using a laser guide star for all observations. We do not consider any fuel limitations or overheads related to the separated LGS spacecraft. While yield studies have shown these are major limitations for a starshade\cite{HabEx_finalReport,Seager2015}\,, here we expect smaller impacts as the LGS may be significantly closer to the telescope, less massive, and thus requires less fuel and time to re-position itself or more than one LGS spacecraft could be deployed. For each guide star magnitude, the raw contrast maps are provided through the results in Sec.~\ref{subsec:Contrast_Performance}. 

The noise floor that defines the minimally detectable planet-to-star flux ratio is not well known. Its value depends on the signal-to-noise of the observation as well as the differential aberrations between the images used for post-processing. In this study we update the calculation of the noise floor in the AYO code. While previous studies adopted a uniform noise floor independent of stellar magnitude, here we adopt a noise floor proportional to the raw contrast. We therefore consider two cases where the 5$\sigma$ noise floor to be equal to a half and a tenth of the raw contrast at each magnitude. These factors correspond to the expected performance of the Roman Space Telescope Coronagraph Instrument with and without noise, respectively, and without a model uncertainty factor in the initial correction of static aberrations\cite{Ygouf2021} (as also assumed in this study). The noise floor is degraded quasi uniformly in the DH due to the initial aberrations and slowly restored depending on the stellar magnitude thanks to the internal AO system. Since EEC around early type stars have the largest flux ratio with respect to their host stars, it means the earliest type stars are removed from the target list without AO and are then recovered gradually with a decreasing magnitude using AO.

\begin{figure}[t]
    \centering
	\includegraphics[width=10cm]{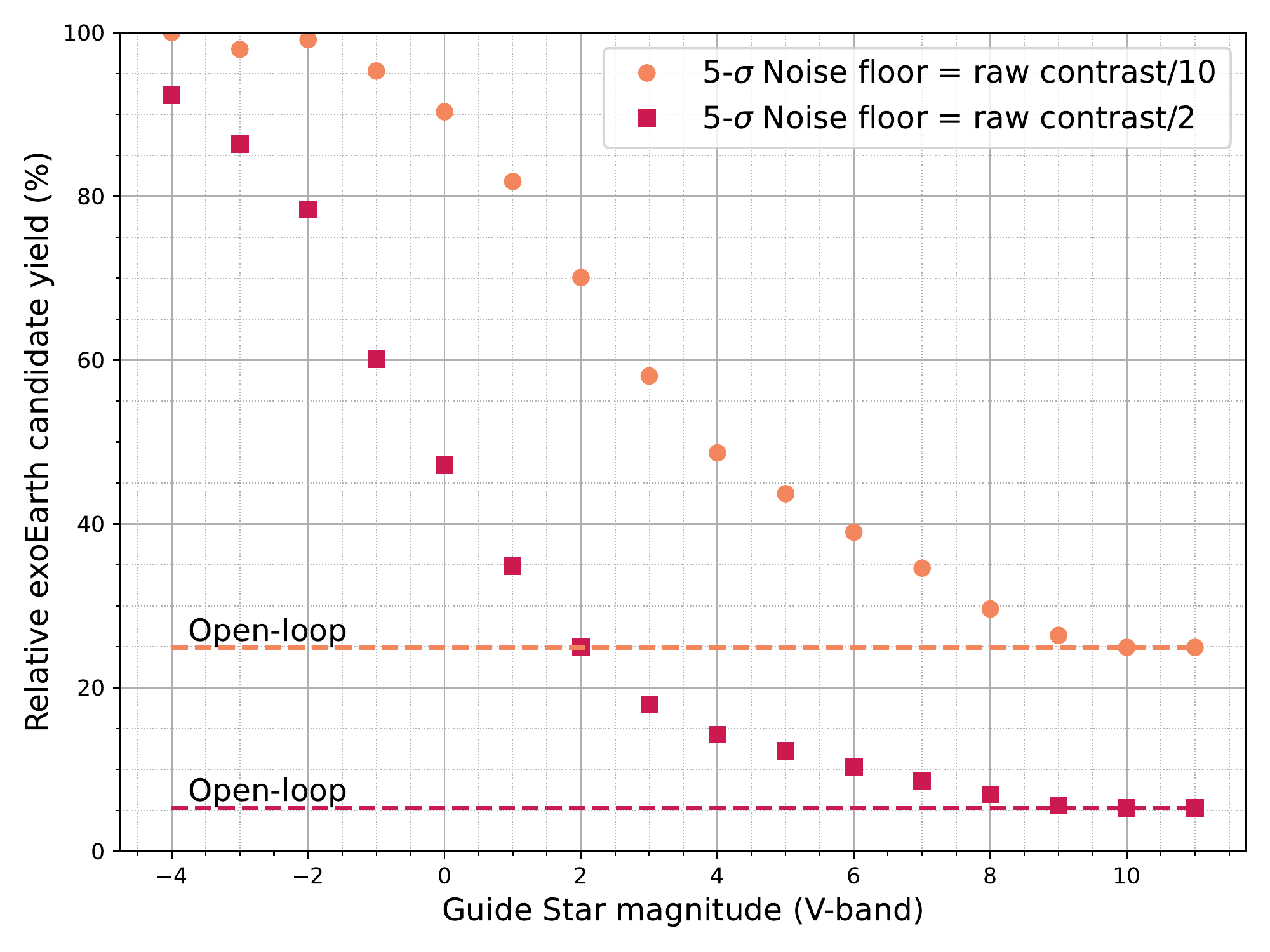}
	\caption{EEC yield calculated with AYO assuming a constant guide star magnitude for all observations. The vertical axis is the relative number of EECs with respect to the case of a perfectly stable telescope. We assume the telescope instabilities follow time series C. In closed-loop, predictive control is performed to correct for the aberrations. Dashed lines represent the performance without the use of AO system. The 5$\sigma$ noise floor is either a tenth (orange, circles) or a half (red, squares) of the raw contrast as calculated in Sec.~\ref{subsec:Contrast_Performance}.} 
	\label{fig:LGS_yield} 
\end{figure}

The AYO results are presented in Fig.\ \ref{fig:LGS_yield}. Without aberrations, 34.1 EECs are detected for the best post-processing scenario and 30.1 in the worst case. We note that these values are considerably lower than the LUVOIR-A yield\cite{LUVOIR_finalReport} because we only model the narrow field-of-view coronagraph designed for LUVOIR-A, while the LUVOIR study used a total of three APLC masks of varying angular scales, and also because of the more pessimistic post-processing noise floor. For the purposes of this work, we therefore emphasize the \textit{relative} yield to estimate the potential gains provided by the AO system. The large difference between the two noise-floor scenarios implies that the yield is mainly limited by the attainable noise floor rather than by the raw contrast. In the case of time series C, 75.1\% and 94.7\% loss in yield is expected without AO for the two post-processing scenarios, respectively. On the other hand, AO associated with a LGS (magnitude $<$ 0) recovers at least 47.2\% of the EECs predicted when the telescope is not affected by dynamic instabilities. A stellar magnitude of -4 allows to recover even more than 92.4\% of EECs. A previous study proposed a 5~Watt LGS spacecraft emitting in visible or near infrared and located at least 40,000~km away from the telescope to simulate starlight with magnitude $<$-4\cite{Douglas2019}\,.

A thorough treatment of the NGS case requires implementing a contrast map for each star in the AYO sample that depends on its magnitude, before overall yield calculation. This modification of the AYO code is currently beyond the scope of this paper and will be presented in future work. In lieu of this, we posit that the LGS scenario at high magnitudes ($3\leq M_V\leq 11$) can inform the NGS scenario. Figure~\ref{fig:LGS_yield} shows that the EEC yield is limited at its best to 18\% and 58\% of the number obtained in the no aberration case when the LGS intensity is limited to NGS-like magnitudes. In the simulation of the LUVOIR-A mission, only ten 3rd magnitude stars are surveyed while the stellar sample includes magnitudes up to 11\cite{LUVOIR_finalReport}\,. To provide an estimated yield with NGS, we therefore weigh each LGS yield by the number of observed stars of same magnitude in the AYO sample. We find the relative yield to be 11.7\% if the 5-$\sigma$ noise floor is the half of the raw contrast. It increases to 42.2\% with the most optimistic noise floor prediction. The mission yield would therefore be multiplied by a factor of 1.7 to 2.2 if the instrument is equipped with an AO system that uses NGS. This result is roughly what is expected for 5th-magnitude stars that dominate the surveyed sample (90 observed stars). 

\subsection{Discussion}

\subsubsection{Optimization of the wavefront sensor spatial and temporal sampling}

\begin{figure}[t] 
    \centering
	\includegraphics[width=10cm]{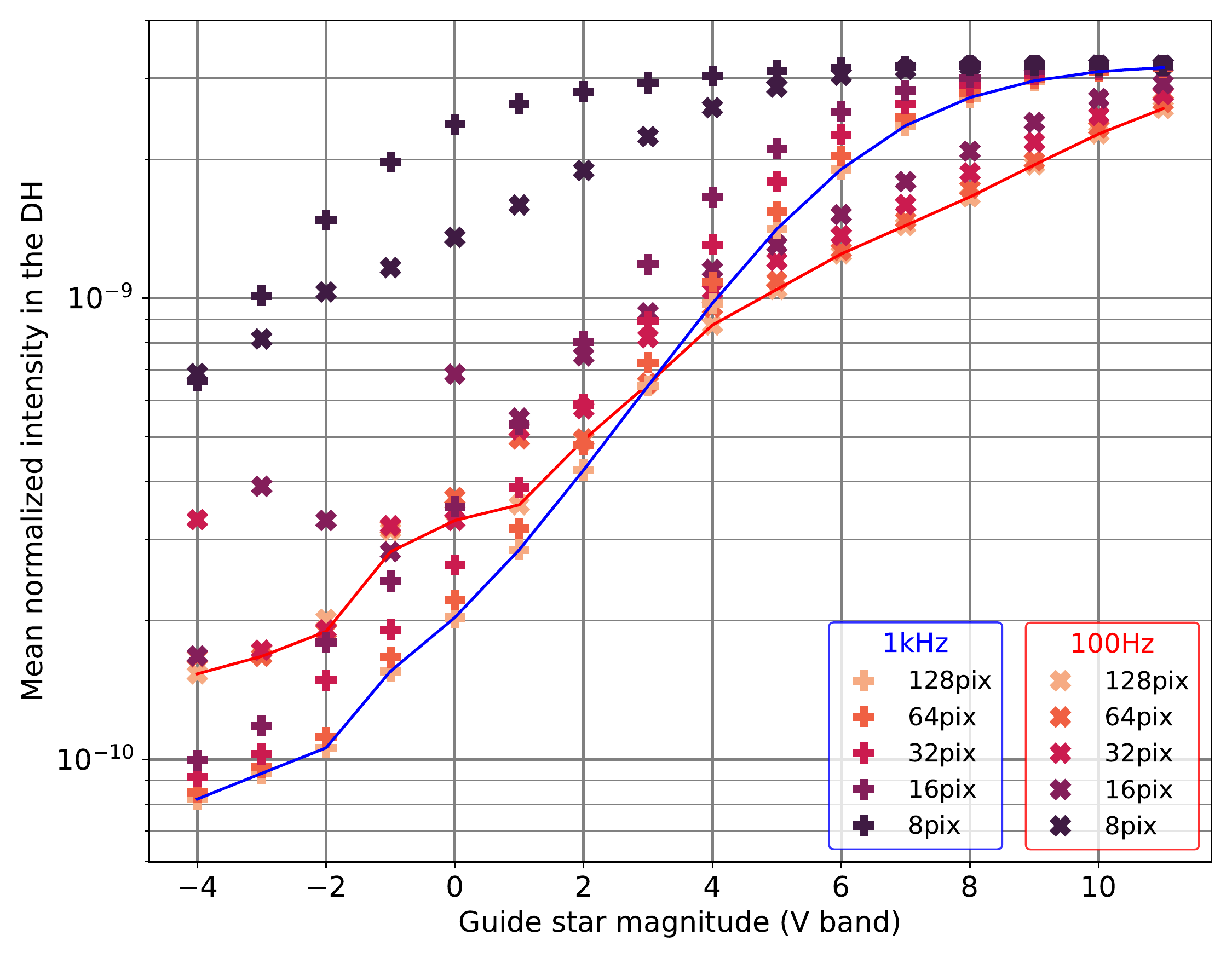}
	\caption{Normalized intensity versus stellar magnitude as a function of the adaptive optics system characteristics. We assume predictive control under the stability conditions of time series C. Red and blue lines represent the best contrast at each stellar magnitude for an AO system running at 100Hz and 1kHz respectively.} 
	\label{fig:Performance_vs_Npix} 
\end{figure}

In Appendix~\ref{sec:WFSphotonnoise}, we discuss how the estimation of modes with significant mid-spatial frequency content could be affected by the spatial sampling of the WFS camera. 
Figure~\ref{fig:Performance_vs_Npix} shows the performance of the AO system with respect to the number of pixels across the beam in the WFS plane. We assume five scenarios where $N_{pix}$~=~8, 16, 32, 64 or 128 pixels across the beam. 
The figure shows that for any observing scenario, the contrast improves with increasing number of pixels used on the WFS camera. 
Indeed, on one hand, the sensitivity of the WFS decreases drastically since it estimates a given mid-spatial frequency aberration with fewer pixels; as a result, $(G^TG)^{-1}$ increases in Eq.~\ref{eq:Mode_PSD_noise} and so does the modal noise.
On the other hand, the relationship between the modal noise and the WFS sampling depends on the noise regime. In the photon-noise limited regime, the total noise of the modal reconstruction is independent of the number of pixels across the beam. In the read-out noise limited regime, the more pixels, the higher is the noise. However, the read-out noise with newer detector technologies is low enough ($RON<1$ electron) that it is negligible with respect to the photon noise, even for stellar magnitude up to 11 where $\sigma_{RON}^2/F_\gamma T=0.06$ for 128 pixels and an exposure time of 10~ms. The WFS is read-out noise limited for stellar magnitudes above 14 with a $RON=1$ electron.

Figure~\ref{fig:Performance_vs_Npix} also shows that a WFS temporal sampling also needs to be optimized.  For natural guide stars with magnitude $M_V\geq3$, the best contrast performance is obtained while running at 100~Hz. For instance, we reach a normalized intensity of $1\times10^{-9}$ with sample C initial perturbation, a stellar magnitude $M_V=5$ and 128 pixels across the beam. For natural guide star where $0<M_V\leq3$, better performance is achieved by increasing the WFS temporal frequency but the resulting contrast remains limited to $>10^{-10}$.
In an LGS scenario ($M_V<0$), an exposure time of $10^{-3}$s is also optimal to reach contrasts down to $10^{-10}$. For such bright stellar magnitude, we did not consider increasing the servo-loop frequency nor the WFS spatial sampling to enhance the contrast performance. Indeed, the performance of the future space-based real time controllers will be limited by the flight-ready computer technology, especially the power consumption, heat dissipation, the size and efficiency of the processors, and the required RAM to make the calculations\cite{Pogorelyuk2021}\,. Future research will determine the highest practical loop frequencies
        
\subsubsection{Random segment phasing error}

\begin{figure}[t]
    \centering
	\includegraphics[width=10cm]{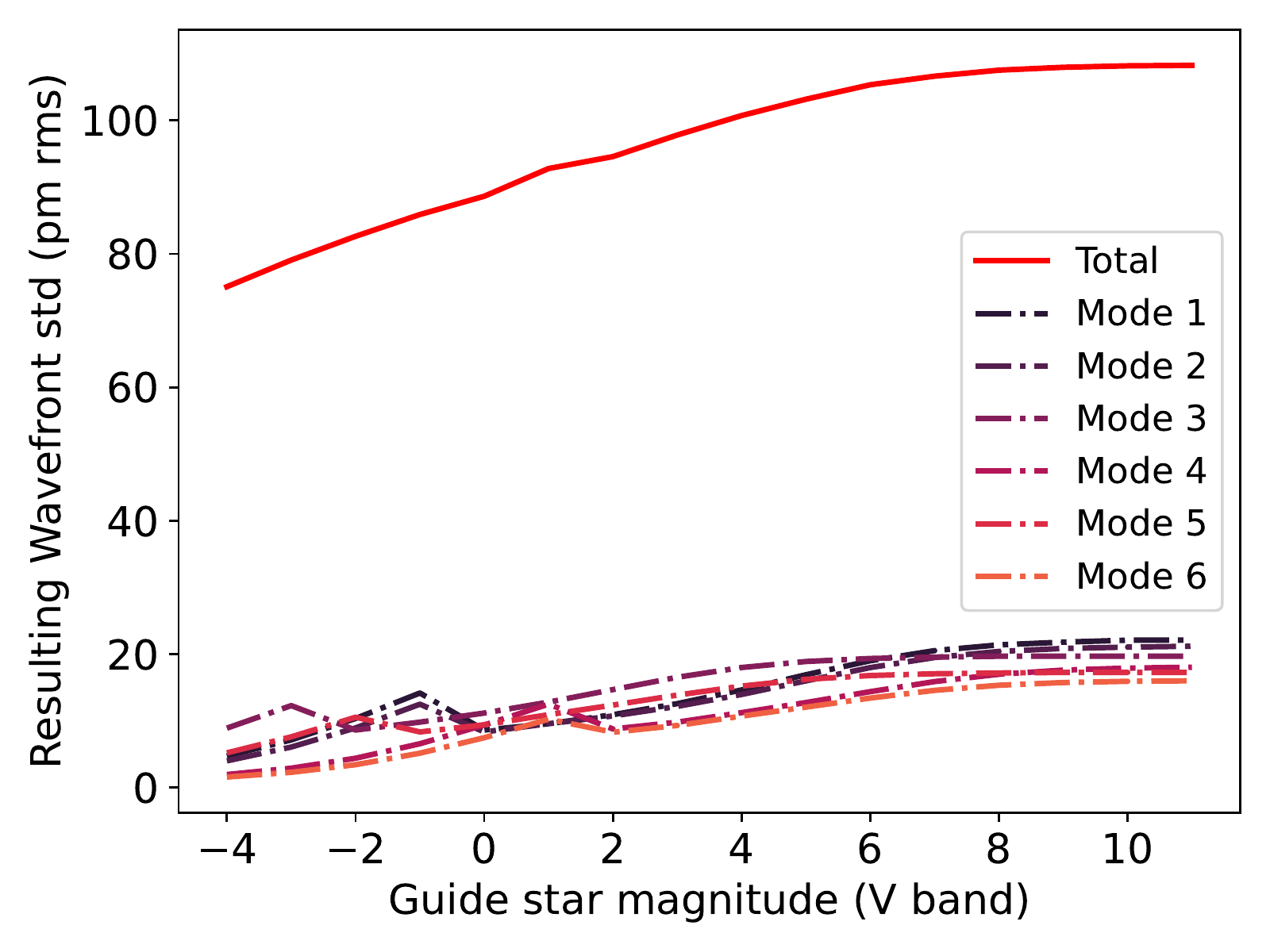}
	\caption{Correction of a random segment phasing error of 110~pm RMS over the pupil. The standard deviation of the residual aberrations is plotted with respect to the stellar magnitude. We assume a predictive control algorithm running at 1000~Hz and a WFS sampling of 64 pixels across the beam.} 
	\label{fig:VariancevsMagnitude_RS} 
\end{figure}

The efficiency of the AO algorithm described above depends strongly on the structural modes in the assumed OPD time series. For the examples above, most of the variance is contained in just a few modes, which assists the WFS measurement thanks to a higher signal-to-noise ratio for these few modes. We can however imagine a worse case where each segment is displaced randomly with respect to its neighbors. For the sake of illustration, we create a new time series where each of the 120 segments has a random piston motion. The total standard deviation is 110~pm RMS (close to sample C for comparison) and each segment carries 1/120th of the total variance. There is no particular vibrations in the segment basis and the temporal PSD is proportional to $f^{-1}$. Following the recipe above to correct for this time series, we first apply PCA, but since each segment movement is random, the variance is distributed among many more modes. The 10 first modes represent only 31.5\% of the total variance. In order to correct for a larger part of the signal in this simulation, we need to consider the 40 first modes that represent 70\% of the total variance. The 80 other modes left over carry individually less than 1\% of the total variance. In this particular time series, the WFS signal per mode is lower than the previous case while the photon noise remains constant and there is no noticeable structure in the principle components. Thus, it is harder for the WFS to distinguish between two separate modes. This induces stronger cross correlations between the modes and higher diagonal terms in the WFS covariance matrix. This can be mitigated with a higher spatial resolution, but at the cost of increasing the computational resources. Figure~\ref{fig:VariancevsMagnitude_RS} shows the standard deviation of the residuals, after the use of a predictive control algorithm running at 1000~Hz, versus the stellar magnitude for such a time series. It shows a dramatic result where the wavefront is not sufficiently corrected, even with the use of a bright LGS, because of the loss of modal signal. This result implies the structures that may be damped by optimizing the design of the telescope are also the easiest modes to correct with the on-board AO system. Therefore, the structures seen in the time series above should be minimized on the sole condition that it does not introduce new modes spread randomly on more segments, even with if the level of optical aberration is globally decreased.

\section{Conclusion}
The success of a future segmented space telescopes designed for the direct imaging of exoplanets with a coronagraph instrument strongly depends on the wavefront stability. However, the mechanical structure and payload of segmented telescopes may introduce quickly varying wavefront aberrations at the coronagraph instrument. We showed that wavefront errors induced by the spacecraft instabilities whose level is $\sim$100~pm~RMS would cause a loss of at least 75\% of the Earth-like exoplanet yield if left uncorrected. We therefore considered a range of possibilities to minimize the aberrations upstream of the coronagraph. First, the telescope dynamic aberrations may be optimized by damping some structural modes, or with the use of primary mirror and telescope control mechanisms, to reach an aberration level of about 10~pm RMS (e.g. sample B). This level should be sufficient to recover more than $\sim$90\% of the yield obtained without dynamical optical aberrations.
If the telescope stability is relaxed such that the dynamic wavefront error entering the coronagraph is $>$10~pm~RMS, the internal coronagraph could be equipped with a high-order AO system to partially correct these errors and stabilize the wavefront. Using NGS, we demonstrated the mission yield would be multiplied by a factor of $\sim2$ with a telescope whose initial stability is 100~pm~RMS (sample C). But the performance is limited by the large exposure time needed for a WFS to achieve picometer-level sensitivity with NGS. A separate out-of-band LGS located in the line of sight may be required to recover 100\% of the yield in a similar configuration. Since such a system might limit the mission yield because of limited fuel and large position displacement overheads, internal laser metrology methods should also be considered. Regardless the initial wavefront error, we find that for the steady conditions potentially offered by the space environment, predictive control algorithms are a valuable tool for correcting OPD errors caused by vibrations.

\appendix    
\section{Statistic properties of an orthonormal basis}
\label{sec:orthonormal_basis}

The following properties are used throughout this paper. Each OPD map $\phi(x,y,t)$ of the time series can be decomposed into a modal basis:
\begin{equation}
\label{eq:modaldecomposition}
\phi(x,y,t)=\sum_{i} a_i(t)Z_i(x,y) \quad ,
\end{equation}
where $Z_i(x,y)$ represents the $i^{th}$ spatial mode of a chosen basis while $a_i(t)$ is its coefficient at a given time. To help the calculations, it is convenient to define the OPD map in an orthonormal basis where, for all modes $i,j$:
\begin{equation}
\label{eq:orthonormalbasis}
\frac{1}{\iint_\text{pupil}\,dx\,dy}\iint_\text{pupil} Z_i(x,y)Z_j(x,y)\,dx\,dy=\delta_{i,j} \quad ,
\end{equation}
where $\delta_{i,j}$ is the Kronecker delta:
\begin{equation}
\delta_{i,j}=\left\{
    \begin{array}{ll}
        1 & \mbox{if} \quad i=j \\
        0 & \mbox{if} \quad i \ne j
    \end{array}
\right. \quad .
\end{equation}
According to Eq.~\ref{eq:modaldecomposition} and \ref{eq:orthonormalbasis}, the modal coefficients can be written as:
\begin{equation}
a_i=\frac{1}{\iint_\text{pupil}\,dx\,dy}\iint_\text{pupil} \phi(x,y,t)Z_i(x,y)\,dx\,dy \quad .
\end{equation}
To calculate the spatial variance of the OPD maps at a given time:
\begin{equation}
\sigma^2(t)=\frac{1}{\iint_\text{pupil}dxdy}\iint_\text{pupil} \phi^2(x,y)\,dx\,dy=\sum_i a_i^2(t) \quad .
\end{equation}
The mean spatial variance over the time series is also easily calculated with an orthonormal basis. Indeed, if $E$ represents the temporal ``expected value", we can write
\begin{equation}
\label{eq:temporal_spatial_variance}
\begin{aligned}
\sigma^2_\phi&=E\big[\sigma^2(t)\big]\\
&=E\left[\sum_i a_i^2(t)\right]\\
&=\sum_i E\big[a_i^2(t)\big]\\
&=\sum_i \sigma_{T,i}^2
\end{aligned}
\end{equation}
due to linearity. The last equation signifies that the averaged spatial variance of the OPD maps $\sigma^2_\phi$ can be calculated as the sum of the temporal variances of the modal coefficients $\sigma_{T,i}^2$. In this paper, the temporal variance of each mode is then calculated as the integral of the $a_i$'s temporal PSD, either pre or post-AO correction.

\section{Long exposure contrast calculation}
\label{sec:long_exp_contrast}
In Sec.~\ref{subsec:Modal_Contrast}, we demonstrated that the total contrast in the DH was dominated by the contrast leakage resulting from the first few PC modes. Also, we illustrated the quadratic relationship between the contrast leakage and the error RMS of these modes. These results allow us to reduce the required computation time to estimate the contrast performance under the different aberration regimes. Indeed, we only need to pre-compute the $\Delta E_i$ due to each mode instead of numerically propagating every single wavefront OPD. At each time step, a reasonable number of modes $n$ (10 modes are sufficient) are scaled by their coefficients $a_i(t)$ and the intensity at the detector is
\begin{equation}
    I(t) = \left|E_0 + \sum_{i=1}^{n} a_i(t) \Delta E_i \right|^2 \quad ,
\end{equation}
where $E_0$ is the diffraction pattern of the coronagraph instrument without any OPD error. The variables $I$, $E_0$, and $\Delta E_i$ are also functions of $x$, $y$, and $\lambda$. Since the frame rate of the science detector is much lower than $\sim$1~Hz, we are ultimately interested in the time average spectral intensity representing long-exposure images and whose expression is given by: 
\begin{equation}
    I = \left\langle\left|E_0 + \sum_{i=1}^{n} a_i(t) \Delta E_i \right|^2\right\rangle \quad .
\end{equation}
where $\langle.\rangle$ represent the temporal mean. This expression can be expanded as
\begin{equation}
\begin{split}
    I &= |E_0|^2\\ 
    &+ 2\sum_{i=1}^{n}\langle a_i\rangle \Re \{E_0^*\Delta E_i\}\\ 
    &+ \sum_{i=1}^{n}\sum_{i^\prime=1}^{n}\langle a_i a_{i^\prime}\rangle\Delta E_i \Delta E_{i^\prime} \quad ,
\end{split}
\end{equation}
where the operator $\Re\{\}$ returns the real part of the field. 
To compute the long exposure images, we make assumptions about coefficient statistics. First, we assume a steady state in the propagation of the aberrations. Therefore, each mode is zero mean: $\langle a_i \rangle=0$. Also, we assume (and have confirmed by calculation) that the modes are independently distributed such that $\langle a_i a_{i^\prime}\rangle=0$ for $i \ne i^\prime$. Under these assumptions, the approximate long exposure image is calculated in the rest of this paper as:
\begin{equation}
\label{eq:MeanContrast_calculation}
    I = |E_0|^2\ + \sum_{i=1}^{n}\langle a_i^2 \rangle|\Delta E_i|^2, 
\end{equation}
where $E_0$ and $\Delta E_i$ are pre-computed while $\langle a_i^2 \rangle$ are equal to the temporal variances of each mode $\sigma^2_{T,i}$ in Eq.~\ref{eq:temporal_spatial_variance}. The stellar intensity at the science detector is therefore a linear combination of $|\Delta E_i|^2$ whose coefficients are the variance of each principal component. This method is similar to the sensitivity analysis method known as the Pair-based Analytical model for Segmented Telescopes Imaging from Space (PASTIS)\cite{Leboulleux2018,Laginja2021}\,.

\section{Spatio-temporal properties of the three OPD time series}
\label{sec:timeseries_properties}
\begin{figure}[t]
    \centering
	\includegraphics[width=\linewidth]{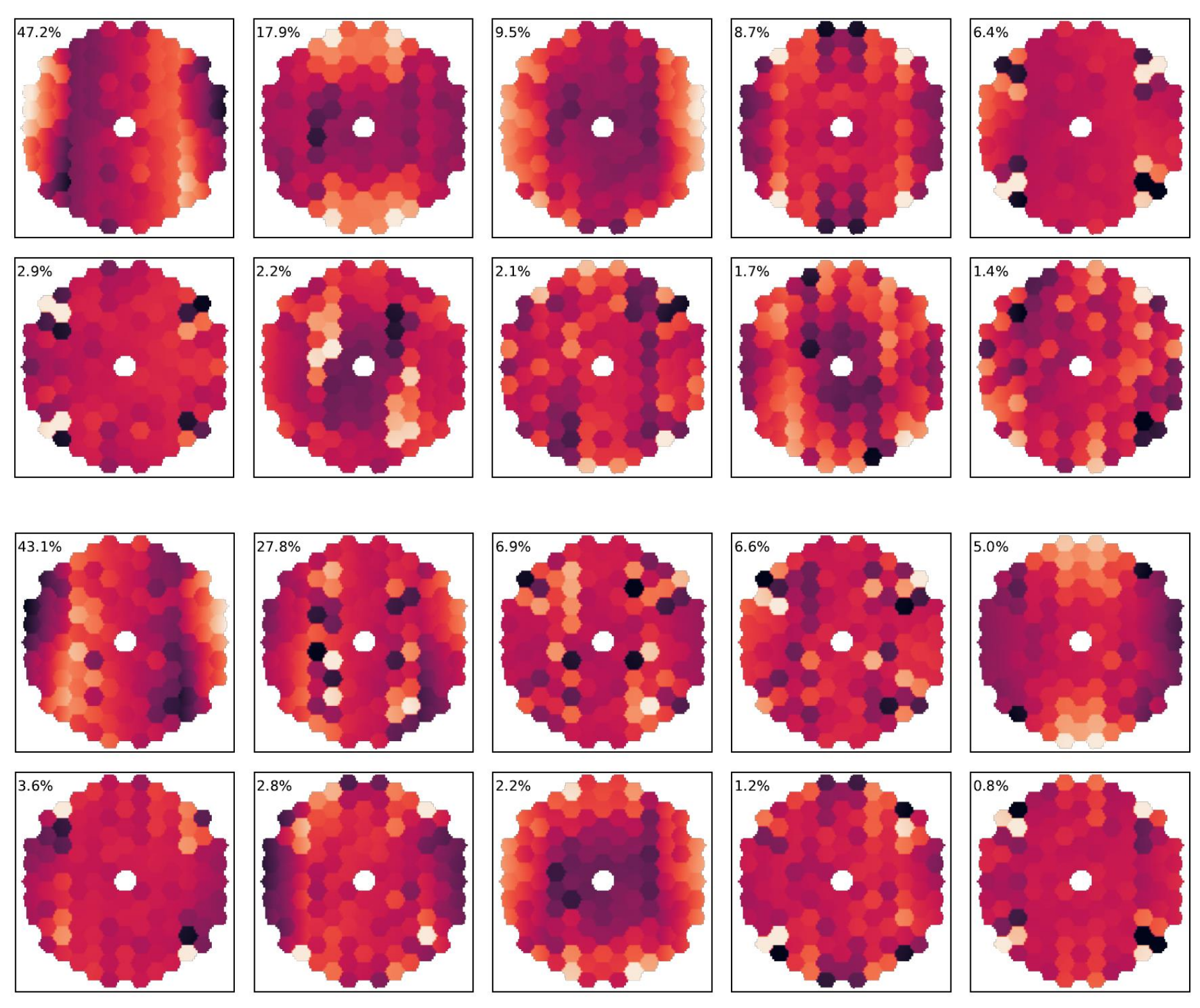}
	\caption{The first ten principal components of the ``sample A" (top) and ``sample C" (bottom) OPD time series. The number in the upper left of each panel is the percentage of the total variance represented by each mode.} 
	\label{fig:PCA_mode_AandC} 
\end{figure}

This section provides the detailed PCA decomposition of time series A and C (also summarized in Table~\ref{table:modeParams}). The standard deviations of samples A and C are respectively $\sim$3.2~pm and $\sim$114~pm. Figure~\ref{fig:PCA_mode_AandC} shows similar spatial modes than sample B, particularly the vertical shape induced by the folding structure of the primary mirror. The total variance is also dominated by a small number of structural modes. For instance the first three modes represent respectively 74.6\% and 77.8\% of the total variance. This decomposition also justifies the use of PCA for time series A and C. Ten modes are corrected in our analysis for sample A while we use 20 modes of sample C for best contrast results at high flux.

\begin{figure}[t]
    \centering
	\includegraphics[width=\linewidth]{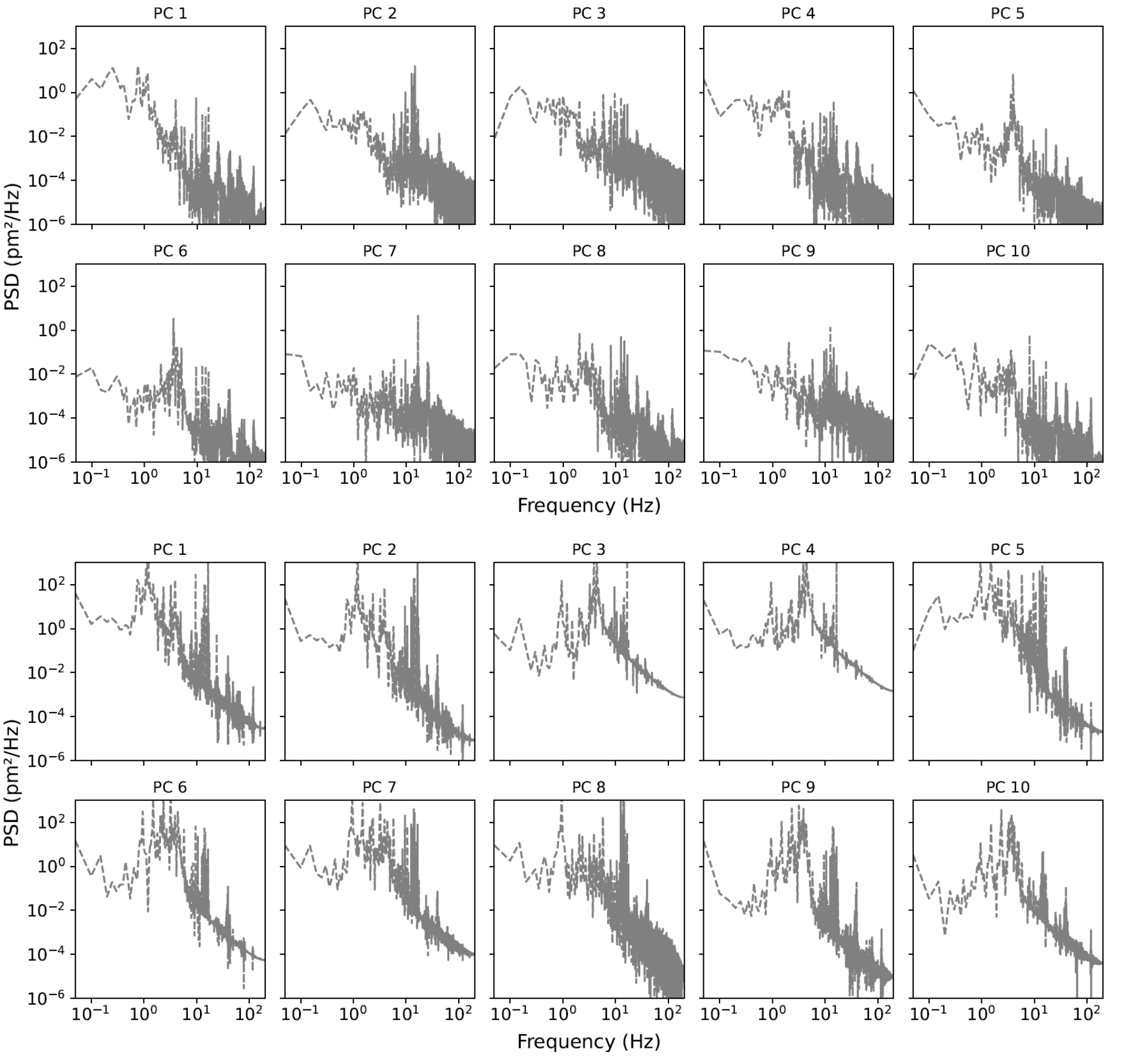}
	\caption{Power spectral densities of the first ten principal components for the ``sample A" (top) and ``sample C" (bottom) OPD time series.} 
	\label{fig:PCAAandC_PSD} 
\end{figure}

The PSDs for each time series are also plotted in Fig.~\ref{fig:PCAAandC_PSD}. The variances are dominated by many localized vibrations for sample C. Most of the energy is contained in vibrations whose temporal frequencies are between 1~Hz and 16~Hz, which are further mitigated with predictive control. However, the vicinity of these vibrations in frequency could limit the performance of predictive control because of the Bode's theorem. On the other hand, sample A's PSDs follow a shape closer to a decreasing power law where an optimized integrator can suffice.

\section{Individual transfer functions of the AO loop elements}
\label{sec:AO_loop_TF}
The total rejection transfer function for one mode is the combination of the transfer functions of every serial element in the AO loop (see Fig.~\ref{fig:AOloop}). Here we consider only the wavefront sensor, the controller, the servo-lag error, and the digital-to-analog converter (DAC). The transfer functions are calculated according to the methods outlined in the abundant literature from the ground-based AO community\cite{Gendron1994,Poyneer2016,Correia2017,Males2018b}\,. To express them, we use the Laplace formalism: $p=2i\pi f$, where $f$ is the temporal frequency.

\subsection{Wavefront Sensor}
The wavefront sensor integrates the incoming perturbation for a period $T$ before passing along the information. Its impulse response is then a gate function of width $T$, corresponding to the servo loop period, centered at $T/2$. We can therefore write its transfer function as a sinc function multiplied by the Laplace transform of a time delay $e^{-i \pi f T}$:
\begin{equation}
H_{WFS}=\frac{1}{2}\text{sinc}(\pi f T)e^{-i \pi f T}=\frac{1-e^{-pT}}{pT} \quad .
\end{equation}

\subsection{Controller}
There are several options for the controller implementation. Here, we compare a simple integrator with a predictive control strategy as done in previous studies in different aberration regimes\cite{Males2018b}\,. We first consider a simple numerical integrator which is the conventional controller in AO. If $M_k$ is the $k^{th}$ WFS measurement, then the $k^{th}$ state for the $i^{th}$ mode, $C_{k,i}$, is
\begin{equation}
C_{k,i}=C_{k-1,i}+g_i M_{k,i} \quad ,
\end{equation}
where $g_i$ is the servo-loop gain for the $i^{th}$ mode. Therefore, since $C_{k-1}$ is the state $C_k$ delayed by the servo-loop period $T$, we can write the relation between the integrator input and its output in Laplace coordinates:
\begin{equation}
C_i=C_ie^{-pT}+g_iM_i \quad ,
\end{equation}
or equivalently,
\begin{equation}
C_i=H_{con,i}M_i \quad ,
\end{equation}
where
\begin{equation}
\label{eq:TF_integrator}
H_{con,i}=\frac{g_i}{1-e^{-pT}} \quad 
\end{equation}
is the controller transfer function.

The ideal rejection transfer function is proportional to the signal-to-noise ratio of the input wavefront as a function of frequency\cite{Dessenne1998}\,. A predictive controller shapes the AO transfer function in this way and aims to catch up with the global time delay introduced by the AO system. This method is particularly efficient for correcting specific vibrations in the OPD time series. The algorithm is based on an autoregressive model where the estimated state $C_{k,i}$ depends linearly on its previous values as well as the previous measurements:
\begin{equation}
C_{k,i}=\sum_{n=1}^{q} b_{n,i} C_{k-n,i}+ \sum_{m=0}^{p} a_{m,i} M_{k-m,i} \quad .
\end{equation}
This approach requires a good model of the open-loop wavefront perturbation to minimize the residuals. Following the calculation described in the case of a simple integrator, we derive the temporal response of such a controller:
\begin{equation}
\label{eq:Predictive_controller}
H_{con,i}=\frac{\sum_{m=0}^{p} a_{m,i}e^{-mpT}}{1-\sum_{n=1}^{q}b_{n,i}e^{-nqT}} \quad .
\end{equation}
This predictive controller reduces to the simple integrator when $p=0$, $q=1$ and $b_1=1$.

\subsection{Servo-lag error}
We have already shown the transfer function of a pure delay. For a delay $\tau$ introduced by numerical calculations and electronic read out time, the servo-lag transfer function is
\begin{equation}
H_\tau=e^{-p\tau} \quad .
\end{equation}

\subsection{Digital to Analog Converter}
The DAC is a zero-order holder whose role is to maintain a voltage on the DM until the voltages are updated by the controller (i.e. it holds the commands during the servo-loop period $T$). Thus, its impulse response is exactly equivalent to the WFS's impulse response and its transfer function can be written as
\begin{equation}
H_{DAC}=\frac{1}{2}\text{sinc}(\pi f T)e^{-\pi f T}=\frac{1-e^{-pT}}{pT} \quad .
\end{equation}

\section{Wavefront sensor sensitivity to random modes}
\label{sec:WFSphotonnoise}
In this section, we describe the propagation of the noise from the detector plane to the modal coefficient estimation to calculate the noise contribution in Eq.~\ref{eq:PSDcorrection}.

\subsection{Qualitative relationship between sensitivity and stellar flux for a zonal reconstruction}
\begin{figure}[t]
    \centering
	\includegraphics[width=\linewidth]{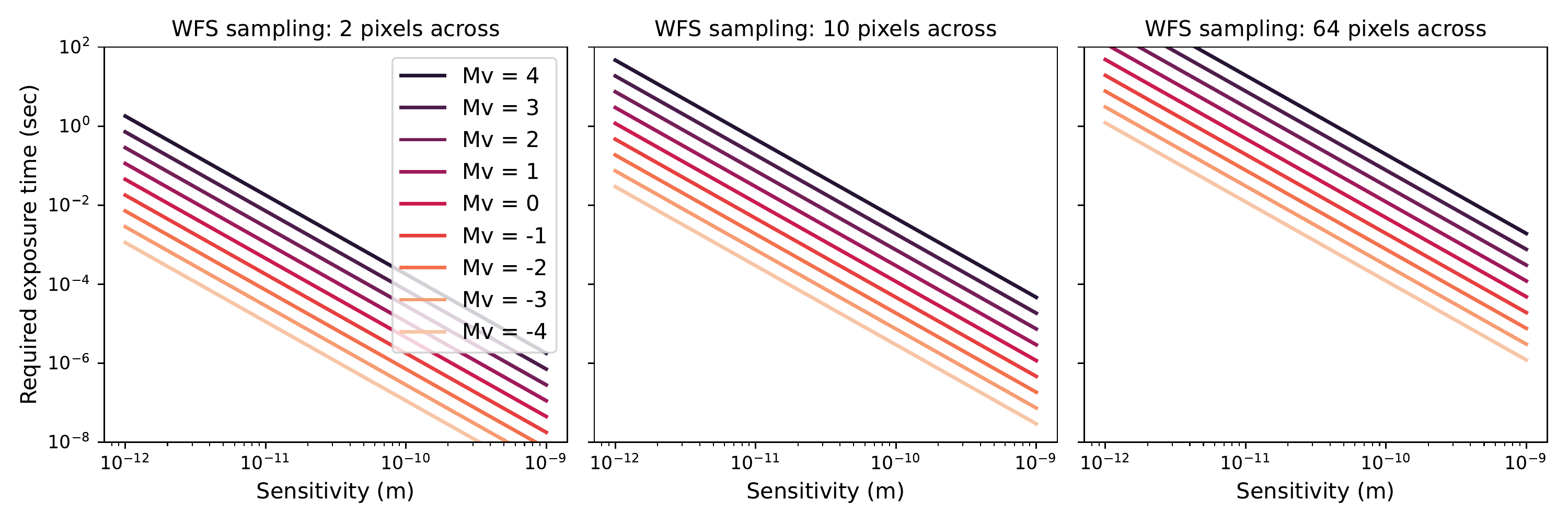} 
	\caption{Exposure time versus individual pixel sensitivity for a ZWFS. We consider 2 (left), 10 (center) and 64 (right) pixels across the pupil.} 
	\label{fig:Exp_timevsSensitivity} 
\end{figure}

First, we describe the impact of photon noise on a wavefront estimation with a ZWFS using a zonal (i.e. local) reconstruction. For a convenient value of $\beta=\sqrt{2}$, the noise level in one pixel of the difference of two WFS images is $1/\sqrt{I_{pix}}$ in radians, where $I_{pix}$ represents the intensity collected by that pixel in a single image in units of photo-electrons. The sensitivity of the WFS then depends on the number of pixels across the pupil, the exposure time, and the magnitude of the observed star. Figure~\ref{fig:Exp_timevsSensitivity} shows the required exposure time for a visible WFS with respect to the desired sensitivity. For example, assuming a goal sensitivity of 1~pm per pixel, a Vega-like star would require an exposure time of 1~sec if the beam is sampled by 10 pixels. However, such a sampling limits the correction to low-order spatial modes. For a sampling of 64 pixels across, we see that almost 100~sec are required for an identical sensitivity and, thus, the correction would be limited to slower drifts and would leave faster wavefront variations uncorrected by the AO system. Another option would be to use laser light, such as with a satellite equipped with a bright laser of negative magnitude to provide more photons\cite{Feinberg2016,Douglas2019}\,. However, zonal wavefront reconstructions are generally less effective than modal reconstructions because they use a small fraction of the starlight to measure the wavefront at each pixel, or zone. We therefore assess the sensitivity of the ZWFS to random orthonormal modes, like in our principal component basis, where all pixels are used.

\subsection{Power spectral density of the noise}
Following previous literature\cite{Males2018b,Douglas2019}\,, we assume the noise on the WFS camera is uncorrelated both spatially and temporally since we only include photon and read out noise. The detector noise can therefore be characterized as white noise with a flat PSD whose amplitude is defined by the variance per pixel $\sigma_n^2$. Even if the light distribution might not be flat on the WFS detector, the incoming noise is assumed identical on all pixels. The signal-to-noise ratio per pixel is therefore
\begin{equation}
\label{eq:SNR-1}
S/N=\frac{F_\gamma T}{\sqrt{F_\gamma T+\sigma_{RON}^2}} \quad ,
\end{equation}
where $F_\gamma$ is the photon rate (photons/sec) collected by each WFS pixels and $\sigma_{RON}$ is the detector read-out noise. Since the Nyquist limit imposes the limits of integration on the PSD between 0 and $1/2T$, we write the temporal PSD measurement noise as
\begin{equation}
\int_0^{1/2T}N_N(f)\cdot df=(S/N)^{-2} \quad ,
\end{equation}
where $N_N$ represents the PSD noise per pixel and finally:
\begin{equation}
\label{eq:Pixel_PSD_noise}
N_N=2(S/N)^{-2}T \quad .
\end{equation}

\subsection{Propagation of noise through the system}
We simulate the propagation of the noise from the wavefront sensor following previous studies in the context of ground-based telescopes\cite{Rigaut1992,Gendron1994,Correia2020}\,. The WFS interaction matrix, $G$, of size $N_{pix}\times N_{modes}$ links the signal received on the WFS camera $s$ to the modal coefficient vector $a$ of the incoming phase:
\begin{equation}
\label{eq:WFS1}
s=G\cdot a \quad .
\end{equation}
The matrix $G$ can be computed analytically thanks to an instrument model or it can be measured after the physical application of each mode in the pupil (with the DM for instance). Then, in order to retrieve the modal coefficient of an incoming wavefront, Eq.~\ref{eq:WFS1} is inverted:
\begin{equation}
\label{eq:WFS2}
\tilde{a}=G^\dagger\cdot \eta \quad ,
\end{equation}
where $\tilde{a}$ stands for the estimate of $a$, $G^\dagger$ is the generalized inverse of $G$:
\begin{equation}
G^\dagger=(G^TG)^{-1}G^T \quad ,
\end{equation}
and $\eta$ is a vector of a pure noise on the WFS camera. The propagated noise covariance $\Sigma_N$ matrix is therefore defined as:
\begin{equation}
\label{eq:Covariance_matrix}
\begin{aligned}
\Sigma_N&=\tilde{a}\cdot\tilde{a}^T\\
&=G^\dagger \eta\,\eta^TG^{\dagger T}\\
&=(G^TG)^{-1}G^TG\left[(G^TG)^{-1}\right]^T\sigma_N^2\\
&=(G^TG)^{-1}\sigma_N^2 \quad .
\end{aligned}
\end{equation}

\begin{figure}[t]
    \centering
	\includegraphics[width=\linewidth]{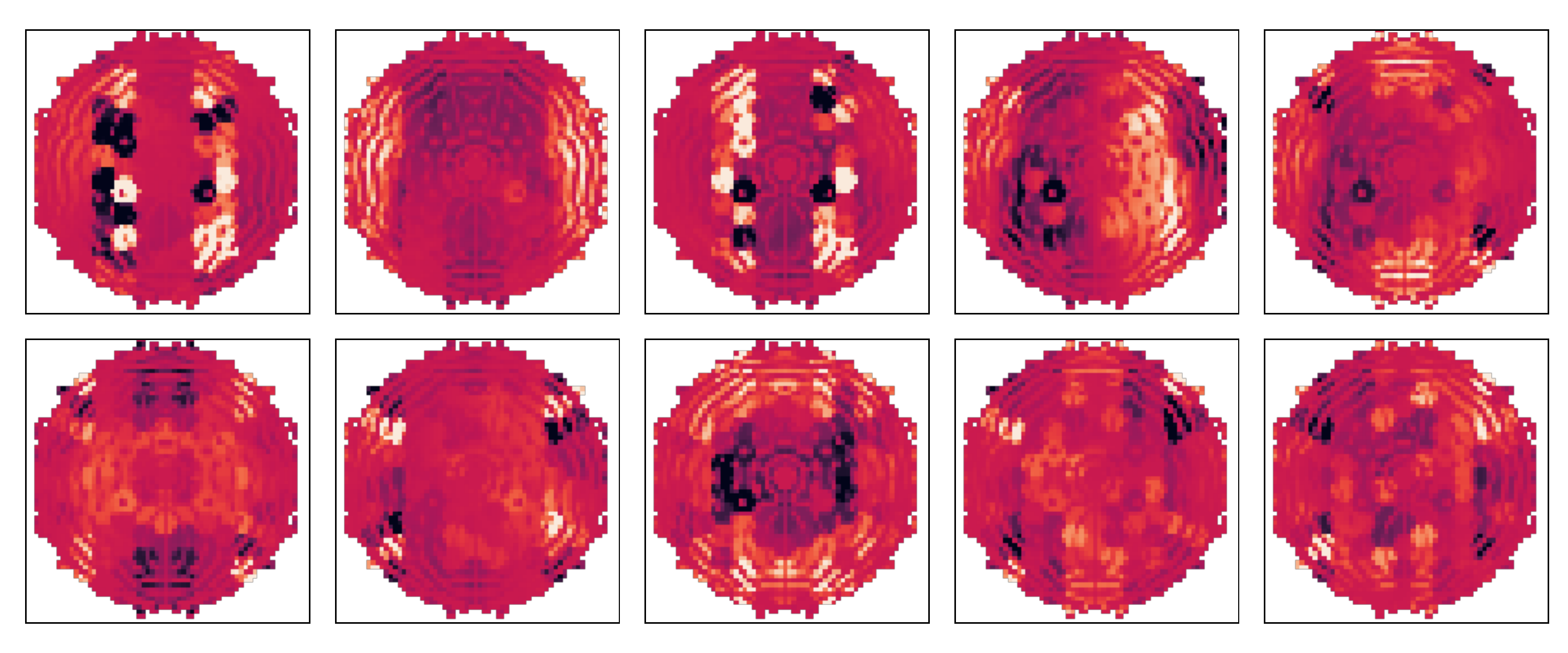}
	\caption{Change in intensity on the ZWFS detector due to each principal components of the OPD time series assuming 64 pixels across the beam and the LUVOIR-A apodized coronagraph, which obscures sections of the pupil.} 
	\label{fig:WFS_images} 
\end{figure}

Finally, the noise PSD for the $i^{th}$ mode $N_i$ is the PSD of the noise reaching the WFS detector and then propagated through the WFS interaction matrix. According to Eq.~\ref{eq:Pixel_PSD_noise} and Eq.~\ref{eq:Covariance_matrix}, this PSD can be calculated as:
\begin{equation}
\label{eq:Mode_PSD_noise}
N_i = (G^TG)^{-1}\Big|_i \cdot 2(S/N)^{-2}T \quad .
\end{equation}
The last equation means that the noise propagation coefficients for each mode are the diagonal terms of $(G^TG)^{-1}$. Since we need to invert the $G$ matrix, modes that are poorly estimated by the WFS result in drastically amplified noise. The number of pixels across the beam in the WFS detector is therefore an important parameter of this study. If the number of pixels in the WFS is too low, the noise propagation per mode increases. However, increasing the number of pixels decreases the signal-to-\ { read-out noise} ratio on the detector. This trade off is studied in more detail in Sec.~\ref{subsec:Contrast_Performance}. Nevertheless, the optimal number of pixels across the beam actually depends on the chosen OPD decomposition. A low-order wavefront sensor with few pixels would be highly sensitive but may be blind to important higher-order aberrations, like segment phasing errors. More pixels would decrease the global sensitivity of the system but allows for correction of mid-spatial frequency perturbations such as those induced by segment phasing errors.

\begin{figure}[t]
    \centering
	\includegraphics[width=9cm]{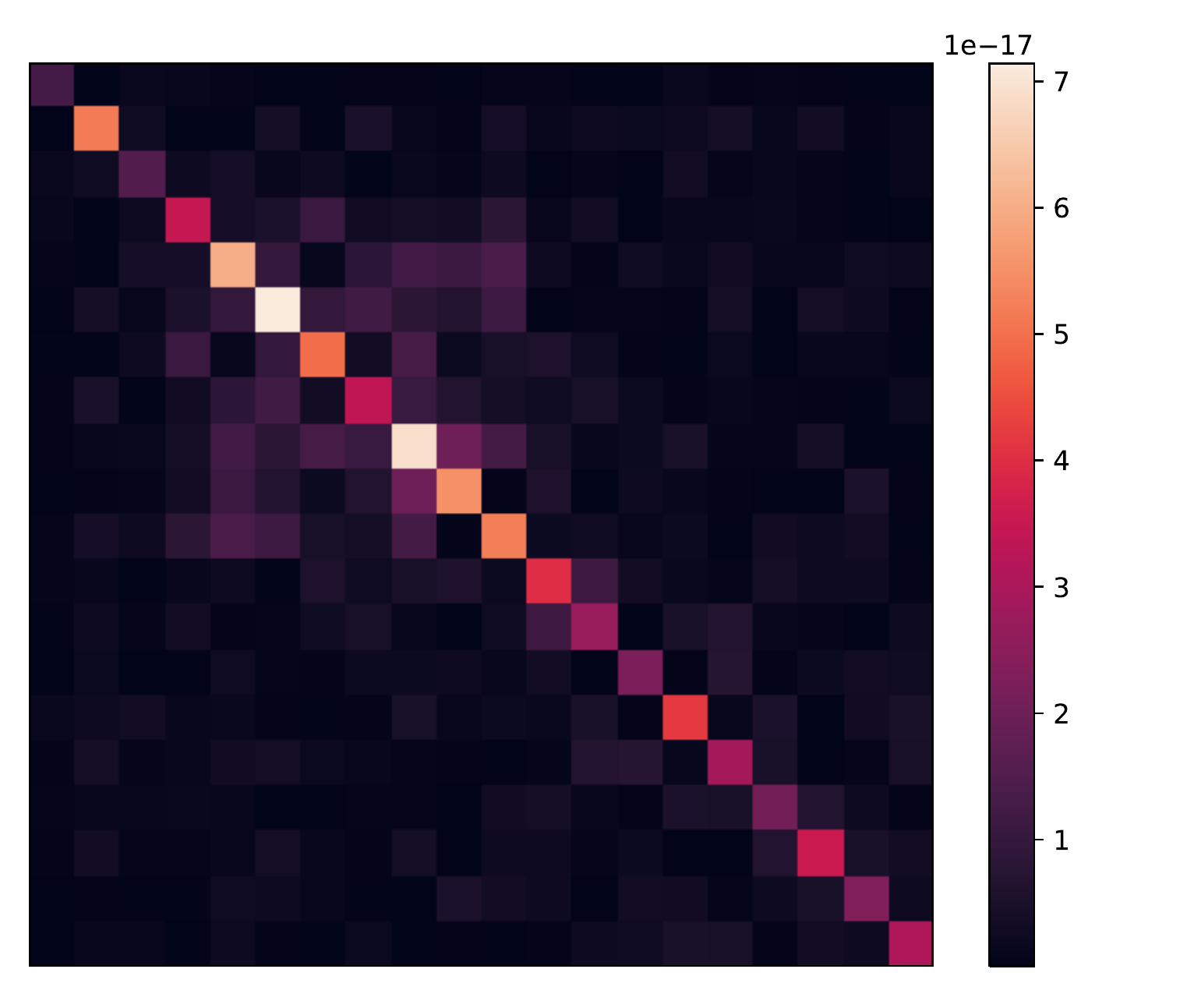}
	\caption{Covariance matrix of the first 20 principal components of time series B as sensed by a ZWFS. In the WFS detector plane, there are 64 pixels across the pupil. The first mode is in the upper left.} 
	\label{fig:Covariance_Matrix} 
\end{figure} 

We illustrate these principles by calculating the matrix $(G^TG)^{-1}$ in Eq.~\ref{eq:Covariance_matrix} for the principle components of time series B, which is also known as the covariance matrix. For a WFS capable of sensing each mode separately, this matrix should be diagonal. To generate it, we use a propagation model of a version of the LUVOIR-A coronagraph instrument equipped with a ZWFS and a WFS camera with 64 pixels across the pupil. The generated WFS images are shown in Fig~\ref{fig:WFS_images} and the covariance matrix is shown in Fig.~\ref{fig:Covariance_Matrix}. In this case, the covariance matrix is not perfectly diagonal and the modes are partially cross-correlated, which means that an individual component injected in the pupil may be confused with other modes. When looking at individual ZWFS images in Fig.~\ref{fig:WFS_images}, the PCA modes are qualitatively recognizable, but the sampling of both the primary mirror and the shaped pupil apodizer has a non-negligible impact on the WFS. Again, this noise can be mitigated by increasing the WFS sampling, but that would result a lower SNR at each pixel for an equivalent stellar magnitude. In this paper, we assume the covariance matrix to be diagonal and we keep the diagonal terms as generated in Fig.~\ref{fig:Covariance_Matrix}.

\acknowledgments 
The authors would like to thank Eric Gendron, Arielle Betrou-Cantou, and Jared Males for fruitful discussions. Part of this work was previously published in "LUVOIR-ECLIPS closed-loop adaptive optics performance and contrast predictions", Proc. SPIE 11823, Techniques and Instrumentation for Detection of Exoplanets X. The research was carried out at the Jet Propulsion Laboratory, California Institute of Technology, under a contract with the National Aeronautics and Space Administration (80NM0018D0004). Integrated Modeling of LUVOIR was performed by Lockheed Martin under a contract with the National Aeronautics and Space Administration (80MSFC20C0017). R. Juanola-Parramon was supported by NASA through the CRESST II Cooperative Agreement 80GSFC17M0002.


\bibliography{bib_GS}   
\bibliographystyle{spiejour}   


\listoffigures

\end{spacing}
\end{document}